\documentclass{article}
\usepackage{graphicx}
\usepackage{indentfirst,csquotes}

\usepackage[utf8]{inputenc}
\usepackage{lipsum}
\usepackage{amsfonts}
\usepackage{amsthm}
\usepackage{algorithm}
\usepackage{algorithmic}
\theoremstyle{plain}
\newtheorem{assumption}{Assumption}
\newtheorem{remark}{Remark}

\usepackage{framed}
\usepackage{tikz}
\usepackage{placeins}
\usepackage[font=footnotesize,labelfont=bf]{caption}
\usepackage{mathabx}
\usepackage{graphicx}
\usepackage{siunitx}

\usepackage{amssymb,amsthm}
\usepackage{xcolor,paralist,hyperref,titlesec,fancyhdr,etoolbox}

\renewcommand\thesection{\arabic{section}}
\renewcommand\thesubsection{\thesection.\arabic{subsection}}

\titleformat{\section}[hang]{\normalfont\huge\bfseries}{\centering\Large\thesection}{10pt}{\Large}

\titleformat{\subsection}[runin]{\normalfont\huge\bfseries\centering}{\centering\large\thesubsection}{10pt}{\large}[.]

\titlespacing*{\section}{0pt}{0ex}{0ex}
\hypersetup{ colorlinks=true, linkcolor=black, filecolor=black, urlcolor=black }

\usepackage{lipsum}

\begin{document}

\title{Optimal Control of Fractional Punishment in Optional Public Goods Game}

\author{J. Grau$^1$ \and R. Botta$^2$ \and C. E. Schaerer$^3$}
\date{\today}
%\address{Address}
%\email{example@mail.com}

\maketitle

\begin{abstract}
Punishment is probably the most frequently used mechanism to increase cooperation in Public Goods Games (PGG); however, it is expensive. To address this problem, this paper introduces an optimal control problem that uses fractional punishment to promote cooperation. We present a series of computational experiments illustrating the effects of single and combined terms of the optimization cost function. In the findings, the optimal controller outperforms the use of constant fractional punishment and gives an insight into the period and size of the penalization to be implemented with respect to the defection in the game.
\end{abstract}

\let\thefootnote\relax
\footnotetext{
$^1$School of Engineering, National University of Asuncion, San Lorenzo, Central, Paraguay -- { E-mail:  jgrau@fiuna.edu.py}}
\footnotetext{
$^2$Polytechnic School, National University of Asuncion, San Lorenzo, Central, Paraguay -- {\tt E-mail:  rbotta@pol.una.py}}
\footnotetext{
$^3$Polytechnic School, National University of Asuncion, San Lorenzo, Central, Paraguay -- {\tt E-mail:  cschaer@pol.una.py}}

%\bigskip
%$\,$
%$\,$

\section{Introduction}

Cooperation is fundamental for success in any human endeavor, but sustaining it is challenging. Over the years, cooperation has been a puzzle studied in various research areas, including natural and social sciences \cite{Hamilton1964,Axelrod1981,ostrom1990governing}. A key issue in studying the evolution of cooperation is identifying the requirements and mechanisms that can sustain cooperation over time \cite{Nowak2006,Sigmund2007,Hardin1968,Trivers1971}.  Two mechanisms have been extensively studied in the literature: a) the application of incentives (such as punishments or rewards) to increase the level of cooperation in the group \cite{Sigmund2007,Yamagishi1986,Sasaki2012,Nowak1992}; and b) the possibility of abstaining from participating, which allows the continuity of cooperation over time \cite{Hauert2002}.

This paper focuses on enhancing cooperation in public goods games (PGG). In this context, punishment serves as a valuable mechanism for improving cooperation, but can be costly. A punishment system requires resources and thus becomes a public good itself \cite{Yamagishi1986}. Traditionally, punishment is implemented in PGG in two ways: as peer punishment or as pool punishment. These approaches differ based on when they are carried out (before or after the game) and the source of funds for sanctions (cooperators' game benefits or a separate pool set aside to penalize defectors) \cite{Sigmund2010,Botta-2024}.

Several methods for enforcing sanctions other than peer or pool punishment have been proposed in recent literature. Some include randomly selecting a group of cooperators to share the cost of sanctioning \cite{Chen2014}, implementing a fixed amount sanction \cite{Dercole2013}, or capping the maximum payment for a defector \cite{Zhang2017}. In this context, a mechanism called fractional punishment, as presented in \cite{g12010017}, reduces the cost of implementing the sanctioning system by punishing only a fraction of defectors in the population while still achieving overall cooperation within the group. This method allows for increased cooperation at a lower cost. However, it does have a drawback in that the proportion of punished free riders remains constant over time without accounting for expected variations based on the results of the applied sanctions.

To minimize the costs of the incentive systems in PGG, some works define an optimal control problem \cite{wang2023optimization,WANG2022_127308,WANG2019_104914}. However, they do not consider free-riders' frequency with respect to the punishment. Consequently, the novelty in this work relies mainly on two points: a) we define an optimal control problem in the fractional punishment sanctioning system to consider the potential variation of the fractional punishment, and b) we introduce an additional term in the optimization objective function to consider the costs associated with sanctioning a single individual.

The results highlight the adaptability of introducing optimal control. Initially, when free-riders are abundant and the cost of punishment is high, a moderate percentage of free-riders are penalized. As the number of free-riders decreases, a more aggressive sanction is used, reducing the cost of punishment. Eventually, the control effort is reduced to a maintenance level, effectively discouraging new sporadic free-riding. This flexible approach significantly reduces the cost of implementing fractional punishment from multiple perspectives making it a versatile option for the final decision-maker.

Finally, this work builds on the idea that the cost of a punishment system is not only a function of how many free-riders are being punished (as discussed when fractional punishment was first introduced) but also of when and how effectively this punishment is being used. In this context, the optimization problem  presented are the first attempt to answer such ideas.

The contents of this paper are organized as follows:  In \S \ref{seccion_PGG} the OPGG and the fractional punishment model are introduced. The optimal control formulation is defined in \S \ref{seccion_control_optimo}. In \S \ref{seccion_resultados}, the simulation results for the functional terms relative importance are exhibited. These results are discussed in \S \ref{seccion_discusion}, as well as a comparison between relevant punishment strategies and the optimal solutions found. Finally, the conclusions of the work are presented in \S \ref{seccion_conclusion}.

%-----------------------------------------------

\section{Control and fractional punishment} \label{seccion_control_theory_FP}

In this section, we define the optional public goods game with fractional punishment in which the optimal control will be applied.

\subsection{Optional public goods game}\label{seccion_PGG}

The public goods game is an economic experiment that can be used to study the evolution of cooperation in a group. The optional version of the game introduces the possibility of rejecting participation and being independent; therefore, the individual must first decide if he will participate and then whether he will contribute (to be a cooperator) or not (to be a free-rider) to the game. The OPGG system is defined as follows. Let $I=[t_0, t_f]$ denote a time interval for the analyzed period, given the state variable $w(t)$, that is comprised of the frequency of cooperators across time $x(t)$, the frequency of free-riders across time $y(t)$ and the frequency of independents (loners) across time $z(t)$, and for each $t\in I$, the state space is ${\cal S}_3 \subset \mathbb{R}^3$ defined as 
\[{\mathcal{S}_3}:= \left \{ [x, y, z]^T \in \mathbb{R}^3: x, y, z \geq 0 \mbox{ and } x+ y + z = 1   \right\}, \]
hence the state variable space is ${\cal Y} = \{ w(t) \in C(I,\mathcal{S}_3): \dot{w} \in C(I, \mathbb{R}^3)\}$. 

Given an initial condition  $w(t_0)=w_0 \in \mathcal{S}_3$ the state equation has the form \cite{g12010017}:
\begin{equation} \label{eq:FPOPGG}
    \begin{array}{lll}
    \dot{x} & = & x ( p_x(w) {-} \bar{p}(w,v)), \\
    \dot{y}  & = & y ( p_y(w,v) {-} \bar{p}(w,v)),\\
    \dot{z}  & = & z ( p_z {-} \bar{p}(w,v)),
    \end{array}
\end{equation}
where $p_i$ is the expected payoff of the $i$th strategy, and $\bar{p}$ is its average given by $\bar{p} = x p_x {+} y p_y {+} z p_z$; $v$ is the distributed control, that belongs to an admissible space ${\cal U} = L^2(I)$ and represents the fractional punishment (see \S \ref{seccion_FPM}). It can be shown that the problem (\ref{eq:FPOPGG}) is well posed (see \cite{g12010017}).

\begin{remark}
The simplex $ {\mathcal{S}_3}$ remains invariant under the flow of Equation (\ref{eq:FPOPGG}) (see \cite{Hofbauer1998,Zeeman1980}); hence, if the state $w(t) \in  {\mathcal{S}_3}$ at $t=t_0$, then $w(t) \in  {\mathcal{S}_3}$ $ \forall  t$, and it is endowed with the standard norm $\| \cdot \| _2$. 
\end{remark}

\begin{remark}
In Expression (\ref{eq:FPOPGG}), the dynamics rely on the expected payoff $p_i$ of the $i$th strategy  with respect to the average population payoff $\bar{p}$ in the sense that, if  $p_i  >  \bar{p}$, then the proportion of the $i$ strategy increases, otherwise it decreases or stays the same. 
\end{remark}

To define the payoff in an OPPG, consider that a group of $n$ individuals are invited to participate in the game. The group is therefore composed by $n_c$ cooperators, $n_d$ defectors, and $n_l$ loners ($n=n_c+n_d+n_l$). Assuming that each contribution is equal and normalized $(c=1)$, the payoff of each corresponding strategy is \cite{Hauertwz2002,Hauert2004}:
 \begin{equation}\label{equ:pagos_ppg}
    p_c=r\frac{n_c}{s}-1, \quad p_d=r\frac{n_c}{s}, \qquad  p_l = \sigma,
 \end{equation}
where $s$ is the number of players in the game ($s=n_c+n_d$), $r$ is the multiplication factor by which the group contribution is multiplied and $\sigma$ is the non-participating individual payoff. The parameters $r$, $\sigma $, and $n$ are considered under the following assumptions:
   \begin{assumption}\label{Assump:1} 
        { The interest rate on the common pool $r$ satisfies $1 < r < n$.  }
    \end{assumption}
	Condition $1 < r$ means that if all individuals cooperate, they are better off than if all defect. Condition $r < n$ means that each individual is better off defecting by itself than cooperating \cite{Hauertwz2002}.  
   \begin{assumption} \label{Assump:2}
     { The payoff $\sigma$ of the loner strategy satisfies $0 < \sigma < r-1$.   }
   \end{assumption}
    This means that a cooperator in a group of cooperators (profiting $(r-1)$ each) is better off than loners that receive $\sigma$, but loners are better off than a defector in a group of defectors, where the payoff is equal to $0$ \cite{Hauertwz2002}.

\subsection{Fractional punishment model}\label{seccion_FPM}

The fractional punishment is a mechanism to reduce the cost of the sanctioning system by punishing only a subset of the free-riders. Following \cite{g12010017}, in this work we consider that only a fraction  $v$ ($0 \leq v \leq 1$) of the defectors will be punished, modifying the payoff presented in (\ref{equ:pagos_ppg}). In particular, we assume that this set of randomly selected defectors will have their corresponding payoff reduced to $0$, while the remaining free-riders will obtain the normal payoff, having an average payoff:
 \begin{equation}
    p_d = (1-v) \left(\frac{r j}{s}\right) {+} \, v \,0.
\end{equation}

 Expected payoffs are used in equations (\ref{eq:FPOPGG}), since a player does not know in advance the composition of the group he is playing with. To this end, observe that the group composition depends on the frequencies of all strategies in the population.

The corresponding expected payoffs considering the fractional punishment for system (\ref{eq:FPOPGG}) 
take the form \cite{g12010017}: 

\begin{eqnarray}\label{equ:pago_cooperadores_final}
    p_x(x, z)& =\mathbb{E}[p_c]= & \, \sigma z^{n {-}1}{+} ra {+} rxb {+}(1{-}r)z^{n {-}1}{-}1,\\
    p_y(x,z,v) & =\mathbb{E}[p_d]= & \sigma z^{n{-}1} {+} \left(1{-}v\right)rxb, \\
    p_z(x,z,v) & =\mathbb{E}[p_l]= &  \sigma.
\end{eqnarray}
where $a:= \frac{1-z^n}{n(1-z)}=\frac{1}{n}\Sigma_{k=0}^{n-1}z^k$ and $b:=\frac{1-a}{1-z}= \frac{1}{n}\Sigma_{k=0}^{n-2}(n-1-k)z^k$. Then it follows that payoffs in expression (\ref{eq:FPOPGG}) are polynomials.
 
\section{Optimal control formulation}\label{seccion_control_optimo}

In this section, we define a problem of minimization with constraints to find the controller $v$ for the OPGG system (\ref{eq:FPOPGG}). The cost function for minimization comprises the weighting of four objectives, each with its significance \cite{Schaerer-Vecpar,Grau-2022}.

Our control effort $v$ corresponds with the fraction of defectors that are punished.
This control belongs to an admissible space ${\cal U} = L^2(t_0,t_f)$, where in our application $v\in [0,1]$. We indicate the dependence of $w$ on $v$ using the notation $w(v)$. Given a target function $w^*$ in $L^2(t_0, t_f)$ and parameters 
$\alpha_i \geq 0$, we shall employ the following cost function, which we associate with the state equation (\ref{eq:FPOPGG}) \cite{Schaerer-Vecpar,etna_vol40_pp36-57, doi:10.1137/080717481}:

\begin{eqnarray}
 J(w(v)), v) & = &  \frac{\alpha _1}{2} \|w(t_f,v)-w^* \|_2^2 + \frac{\alpha_2}{2}\int _{t_0}^{t_f} \|w(\tau,v)-w^* \|_2^2 d\tau \nonumber\\
 & & + \frac{\alpha _3}{2}\int _{t_0}^{t_f} v^2(\tau) d\tau + \frac{\alpha _4}{2}\int _{t_0}^{t_f} (v(\tau)y(\tau))^2 d\tau. \label{cost:func}
\end{eqnarray}

The weights establish the importance of each of the functional terms, {\it i.e.}, $\alpha_1$ weights the cost of not being in the desired state $w^*$ at $t_f$, $\alpha _2$ weights the accumulated squared error of trajectory $w$,  $\alpha _3$ weights the value of the controller at all times, and finally $\alpha _4$ weights the frequency of sanction individuals at all times. In other words, the term associated with $\alpha _4$ measures the  the costs of implementing the control effort given the current frequency of defectors in the population, assuming this two values are proportional (as they usually are).

The optimal control problem for system (\ref{eq:FPOPGG}) consists of finding a controller $u \in {U}$ which minimizes the cost function (\ref{cost:func}) as  

\begin{equation}
 {J}(w,u) := \min _{v\in {U}} J(w(v),v)\ s.t.\  (\ref{eq:FPOPGG}).
\end{equation}

\begin{remark}
The integrands of the cost-function {\rm (\ref{cost:func})} belong to $C^2$. Additionally, its associated Hamiltonian function is given by 
\[
    H(t,w,v,\lambda)= 
    \frac{\alpha_2}{2} \|w(t,v)-w^* \|_2^2 +
    \frac{\alpha_3}{2} v^2(t) + 
    \frac{\alpha_4}{2} v^2(t)y^2(t) +
    \lambda^Tf(t,w,v), 
\]
is strictly convex with respect to $v$ for $w_0$ being interior to $\mathcal{S}_3$ and $\alpha_3>0$ or $\alpha_4>0$; then  
\[H_{vv}= \alpha_3+\alpha_4 y^2>0, \forall t\geq t_0.\] This determines necessary and sufficient conditions to obtain a unique optimal solution that  minimizes {\rm (\ref{cost:func})} {\rm (}see {\rm \cite{Leitao2014, Speyer-2010}}{\rm )}.

\end{remark}

Next, we discuss the implemented algorithm and the Python's library used for the implementation.

\subsection{The GEKKO library and the algorithm}

We use a Python package called GEKKO for numerical implementation, specifically designed for optimization and control problems \cite{beal2018gekko}.
This package offers three free solvers in its public distribution: APOPT, BPOPT, and IPOPT. This work will use the default solver APOPT unless otherwise indicated.

A pseudo-code of the algorithm implemented in the GEKKO is presented in Algorithm \ref{Geko-Alg}.
    
\begin{algorithm}[thb]
    \caption{Solving the minimization of the cost function. \label{Geko-Alg}}
    \begin{algorithmic}[1]
        \STATE Declare initial parameters such as number of points, initial and final time, maximum value for $v$, initial state $w_0$, weights $[\alpha_1,\alpha_2,\alpha_3,\alpha_4]$, and the values of $n$, $r$ and $\sigma$ mentioned in section \ref{seccion_control_theory_FP};
        \STATE Initialize the model;
        \STATE Declare all variables and equations;
        \STATE Define the cost function as a minimization objective;
        \STATE Choose a solver and execute it;
        \STATE Display results;
    \end{algorithmic}
\end{algorithm} 
\FloatBarrier

\section{Results}\label{seccion_resultados}

To appropriately choose the weights in the cost function (\ref{cost:func}), we start by analyzing each element on its own, for example, $[\alpha_1,\alpha_2,\alpha_3,\alpha_4]=[1,0,0,0]$, to understand the single importance of $\alpha_1$ and so forth. Then, we proceed to examine the most important pairs of combinations between the elements of the cost function (such as $\alpha_2$ and $\alpha_3$ or $\alpha_2$ and $\alpha_4$). Lastly, we merge the previous findings to obtain a combination that yields better results, which we compare with relevant constant fractional punishment strategies.

In all cases, the parameters mentioned in \S \ref{seccion_control_theory_FP} necessary for defining the problem are as follows: $n=5$, $r=3$, and $\sigma=1$. To simplify the notation, from now on, we will refer to the importance of $\frac{\alpha _1}{2} \|w(t_f,v)-w^* \|_2^2 $ as ``the importance of $\alpha _1$". Similar acronyms will be used for the other terms of the functional (\ref{cost:func}).

\subsection{Importance of $\alpha _1$} \label{subS_a1}

\begin{figure}[htb]
    %\begin{center}
    %\begin{tabular}{ccc}
    %\includegraphics*[width=0.5\linewidth]{Img/SS_1000_600dpi.pdf} \\ a)
    %\end{tabular}
    %\begin{tabular}{ccc}
    %\includegraphics*[width=0.4\linewidth]{Img/vT_1000_600dpi.pdf} & %\includegraphics*[width=0.4\linewidth]{Img/yvT_1000_600dpi.pdf} \\
    % b) & c)
    %\end{tabular} 
    %end{center}
    \begin{center}
    \begin{tabular}{cc}
    \includegraphics*[width=0.45\linewidth]{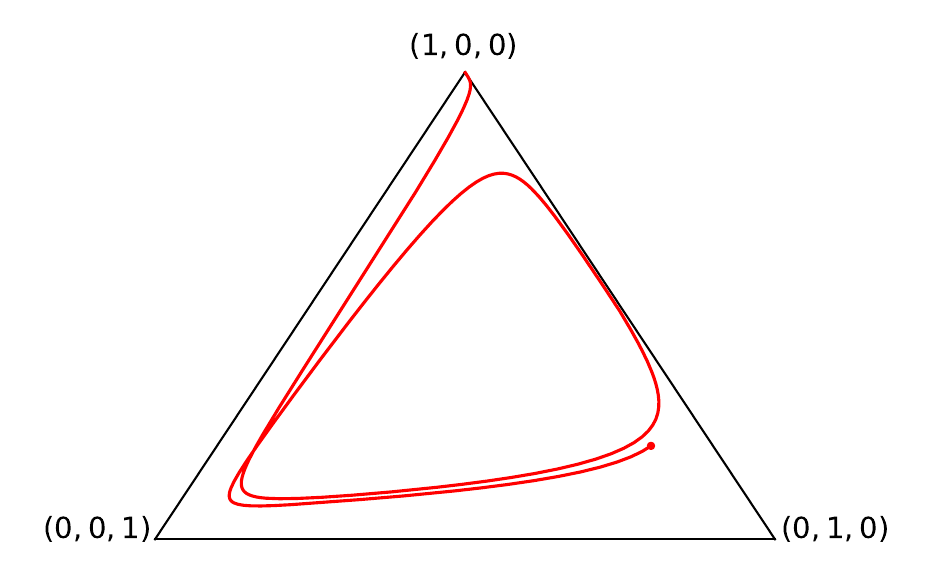} & 
    \includegraphics*[width=0.4\linewidth]{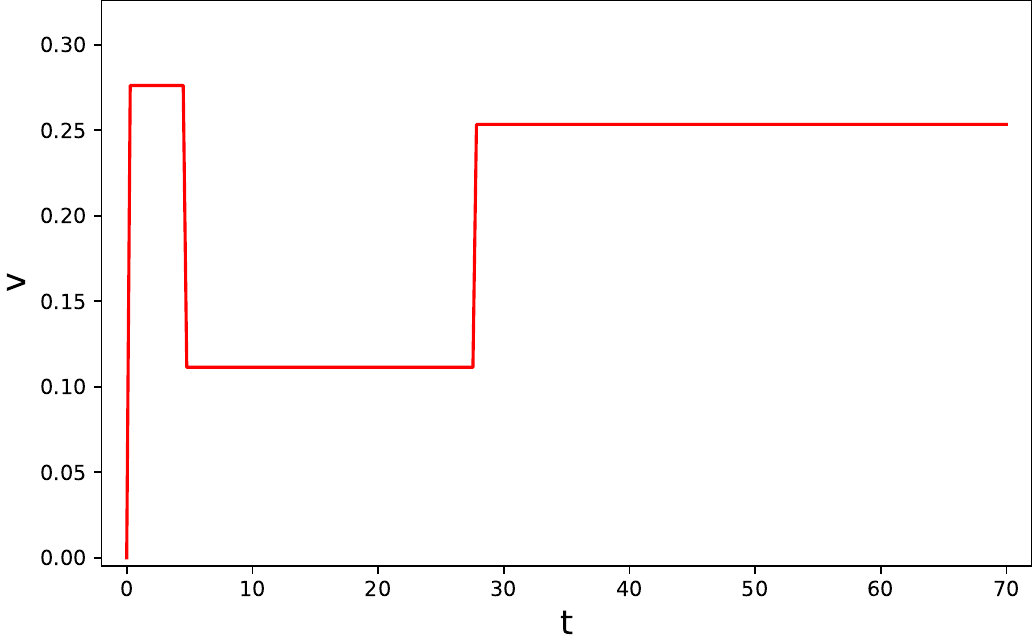} \\
     a) & b)
    \end{tabular} 
    \end{center}
    \caption{{\bf Importance of $\alpha _1$. Part 1.} Results obtained by simulating from 0 to 70 time units with 250 steps, from initial state $w_0=[0.2,0.7,0.1]^T$ (marked with a dot), using the controller curve shown in b). a) State space and trajectory of the system. b) Control effort $v$ over time.}
    \label{fig:a1}
\end{figure}

The value of $\alpha_1$ is important for minimizing the final state error at time $t_f$. This situation is observed in Figure \ref{fig:a1}, which illustrates a simulation where the final state is close to full cooperation. However, when the only requirement is for the final state to be close to full cooperation (vertex $x=1$), the trajectory can oscillate in the state space and closely approach the vertex $z=1$. 
This situation means that in the context of a public good, simply minimizing this error does not guarantee sustainability for the whole period since approaching $z=1$ could lead to bankruptcy caused by the abandonment of most of the game participants.

%\FloatBarrier
\begin{figure}[htb]
    \centering
    \includegraphics*[width=0.4\linewidth]{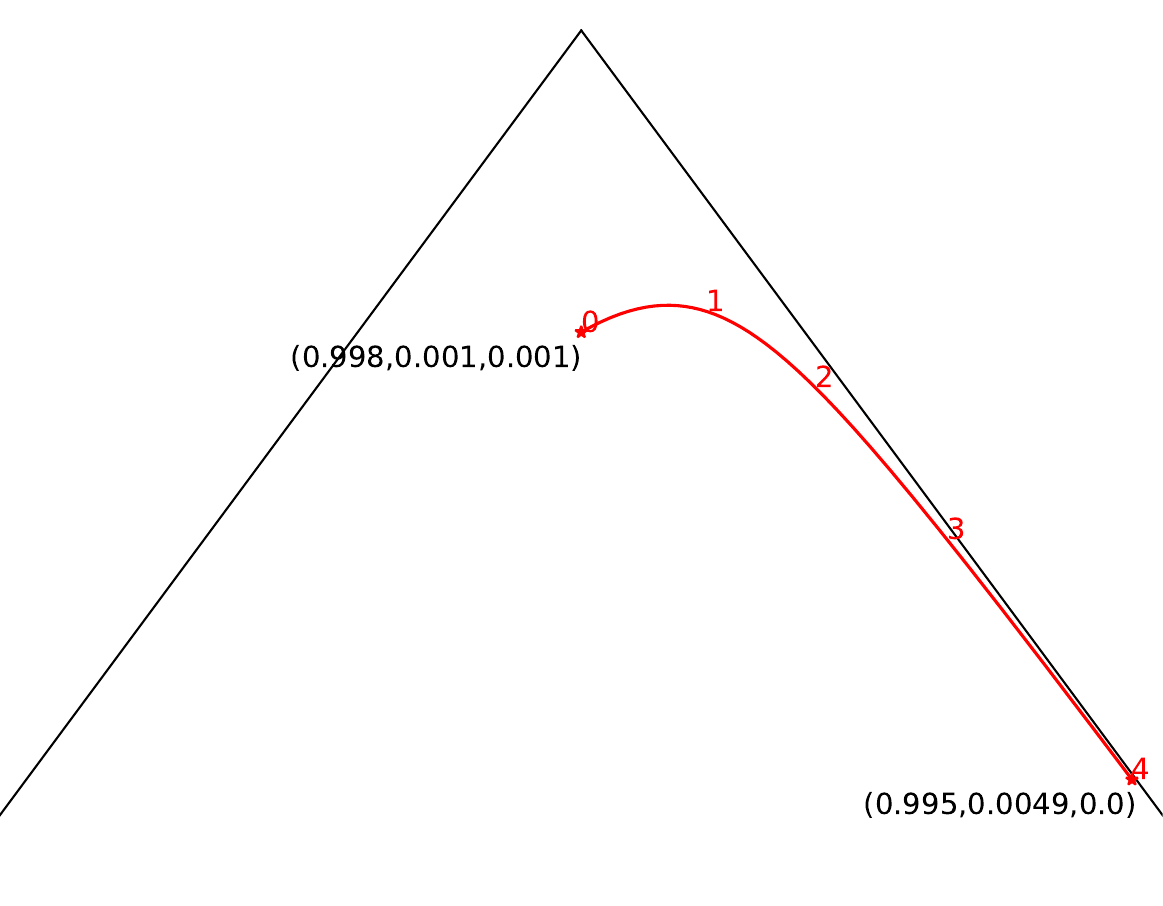}
    \caption{{\bf Importance of $\alpha _1$. Part 2.} Result obtained by simulating from 0 to 4 time units with 600 steps, with no controller ($v=0$), starting at $w_0=[0.998,0.001,0.001]^T$. A zoomed in simplex border can be seen in black, along with red time stamps in the trajectory traveled. The initial state error amounts to around 0.002449 while the final state error amounts to 0.007008, showing how unnoticeable small time deviations could be at the end of the simulation for this system.}
    \label{fig:zoom1}
\end{figure}
%\FloatBarrier

Note that because the system is continuous, it cannot  quickly deviate from the state of full cooperation (vertex $x=1$).  Hence, if the goal is to minimize the state error at all times (see \S \ref{seccion_resultados_a2}) rather than just minimizing the final state error at time $t_f$, a boundary layer will emerge. This situation is shown in  Figure \ref{fig:zoom1}. However, depending on the desired accuracy, this solution can be acceptable. With this consideration, $\alpha_1$ is excluded from the subsequent analysis by setting $\alpha_1=0$.

\subsection{Importance of $\alpha _2$} \label{seccion_resultados_a2}

\begin{figure}[htb]
    \begin{center}
    \begin{tabular}{ccc}
    \includegraphics*[width=0.6\linewidth]{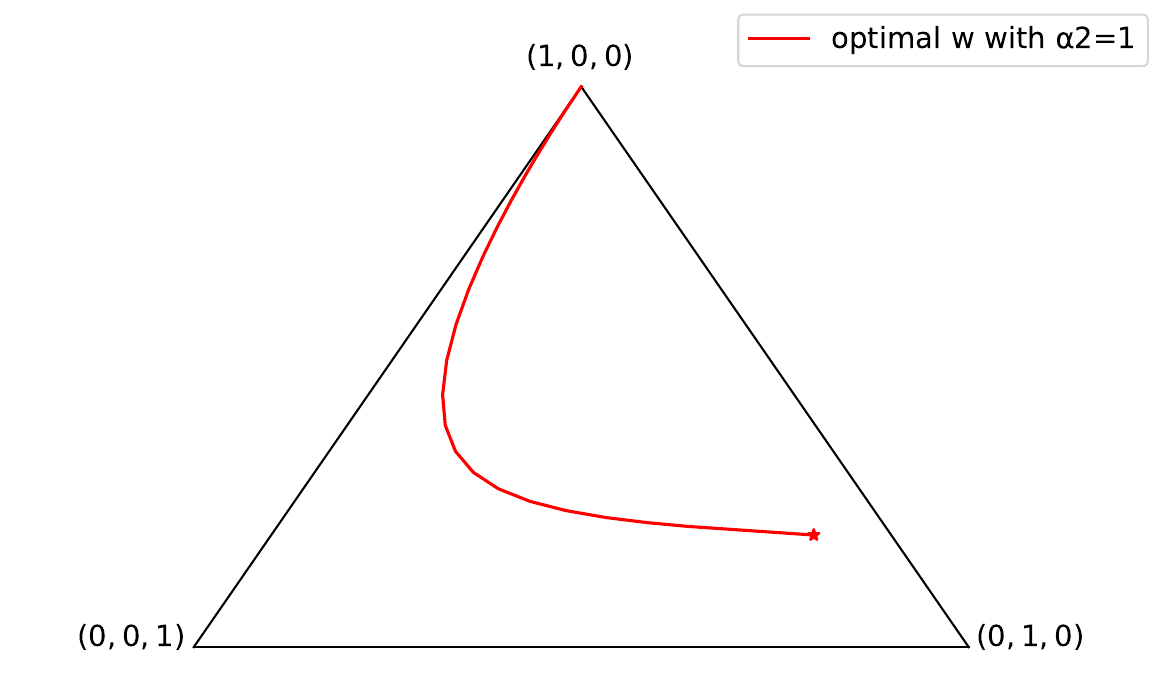} \\ a)
    \end{tabular}
    \begin{tabular}{ccc}
    \includegraphics*[width=0.4\linewidth]{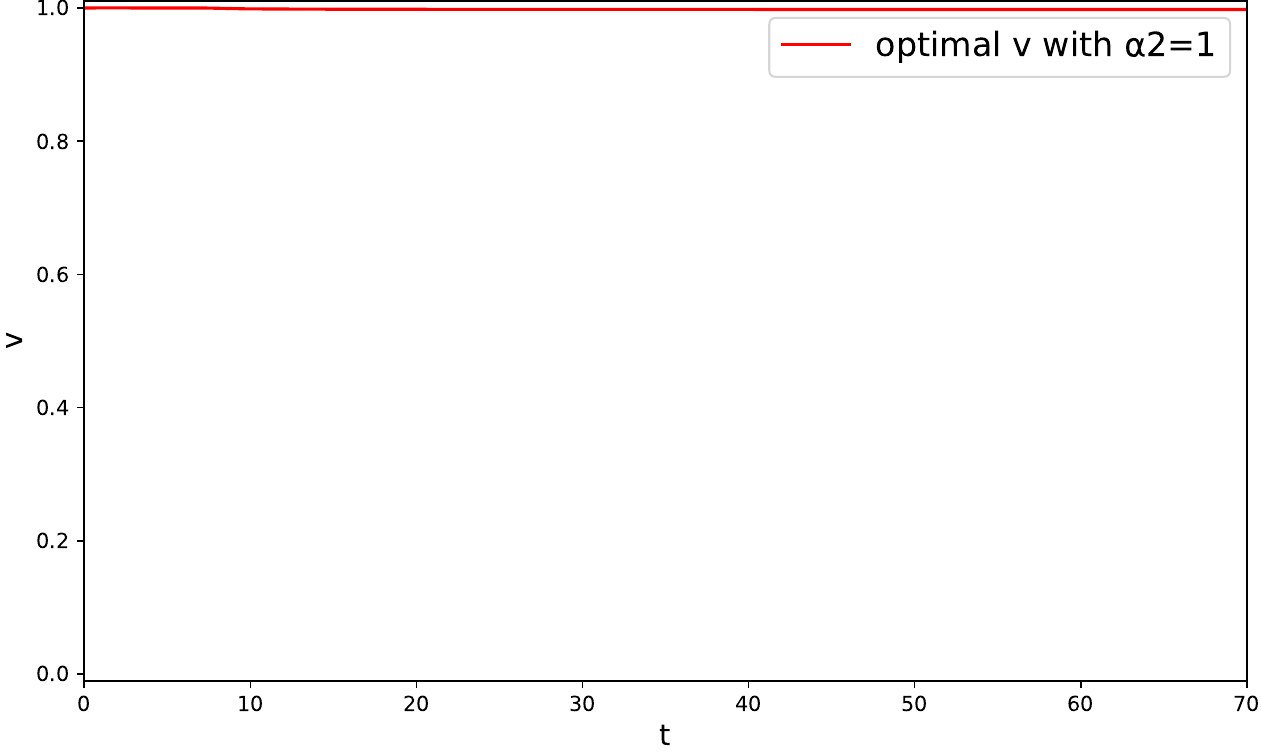} & \includegraphics*[width=0.4\linewidth]{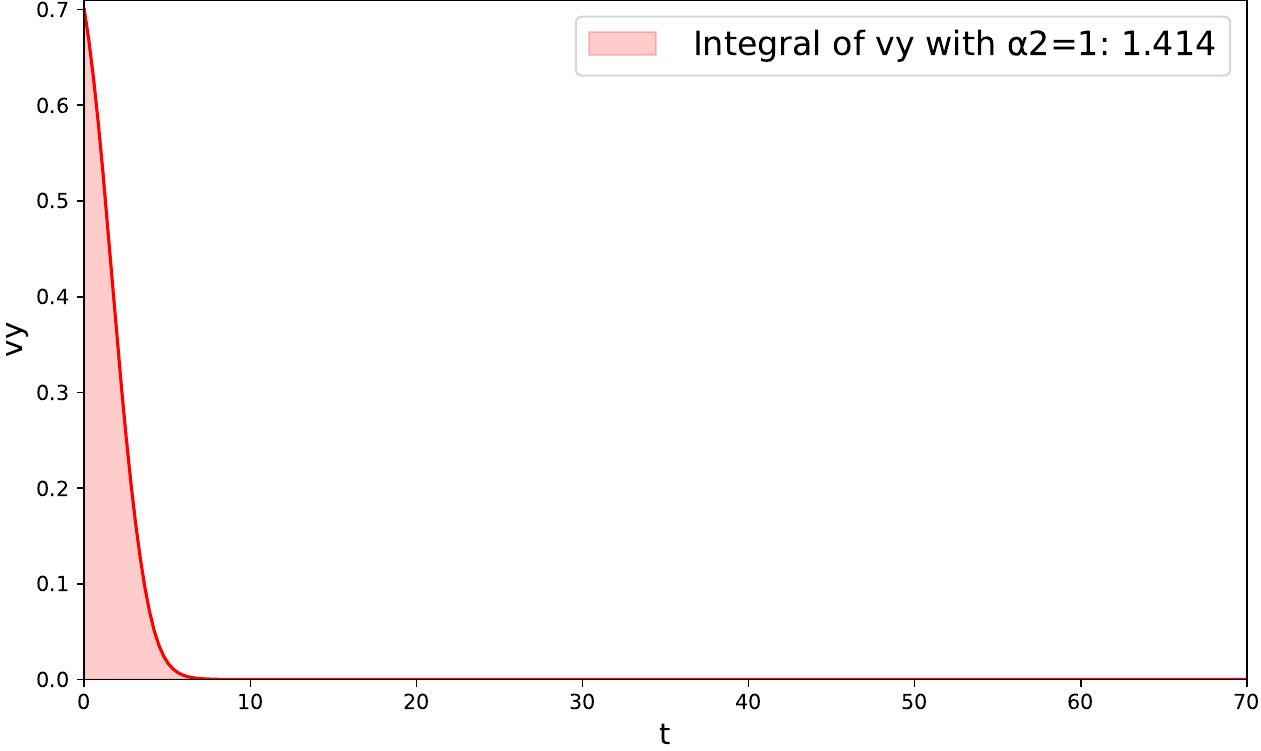} \\
     b) & c)
    \end{tabular} 
    \end{center}
    \caption{{\bf Importance of $\alpha _2$.} Result obtained by simulating from 0 to 70 time units with 250 steps, only considering $\alpha_2$  and with starting state $w_0=[0.2,0.7,0.1]^T$. a) State space and trajectory of the system. b) Control effort $v$ over time. c) Proportion of punished individuals $yv$ over time. The red shade represents the area under the curve and the legend shows its value.}
    \label{fig:a2}
\end{figure}

Considering only the accumulated state error across the entire time period, we aim to reduce the error as quickly as possible without considering the magnitude of the punishment costs. This is observed in Figure \ref{fig:a2}. The control effort reaches its maximum value, indicating that we are penalizing all free-riders, leading to the fastest reduction in error possible. This trend continues throughout the entire period, ensuring that no further error accumulates once we are close to, or have reached, the desired state $w^*$.

%\FloatBarrier

In practice, for most real-world systems, it is impossible to sanction all free riders at all times. This impossibility may be due to the high cost of punishing or the difficulty of identifying and addressing all (or most) free riders. However, the obtained numerical result is the minimum estimation for all scenarios of the accumulated errors.

\subsection{Importance of $\alpha _3$ and $\alpha_4$}

When analyzing the importance of $\alpha_3$, the optimal controller minimizes $v$ to its minimum value ({\it i.e.}, keeping it at zero the entire period). If this occurs, full cooperation is not achieved, defeating the purpose of the whole controller.

\begin{figure}[htb]
    \begin{center}
    \begin{tabular}{ccc}
    \includegraphics*[width=0.6\linewidth]{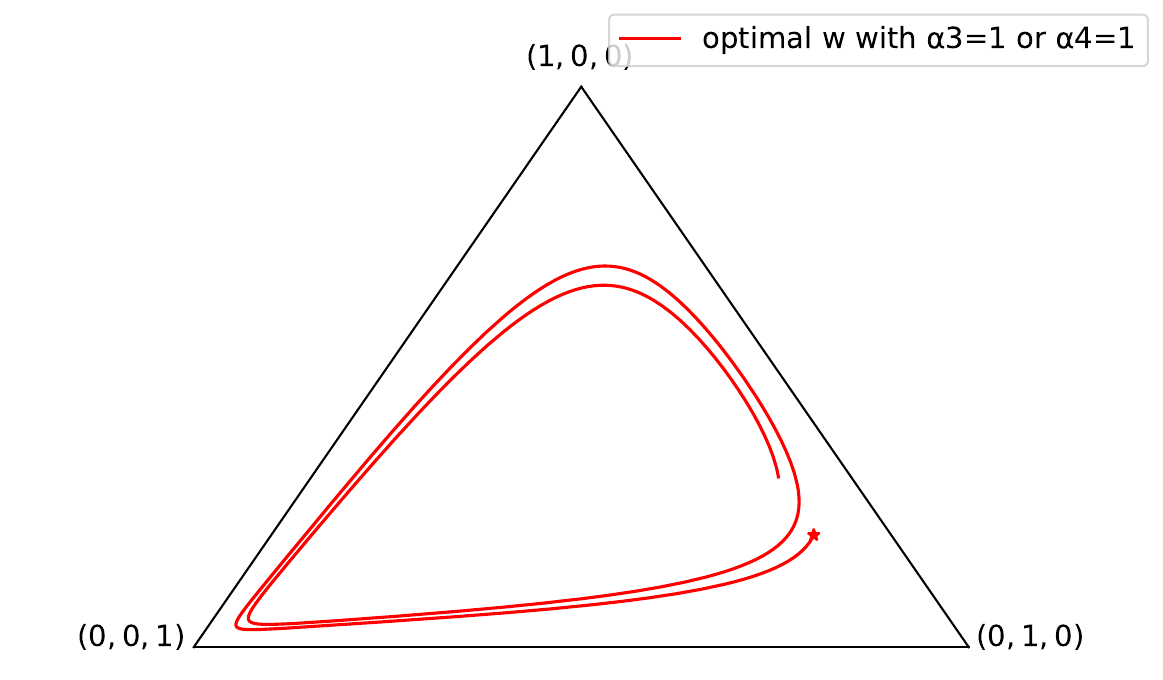} \\ a)
    \end{tabular}
    \begin{tabular}{ccc}
    \includegraphics*[width=0.4\linewidth]{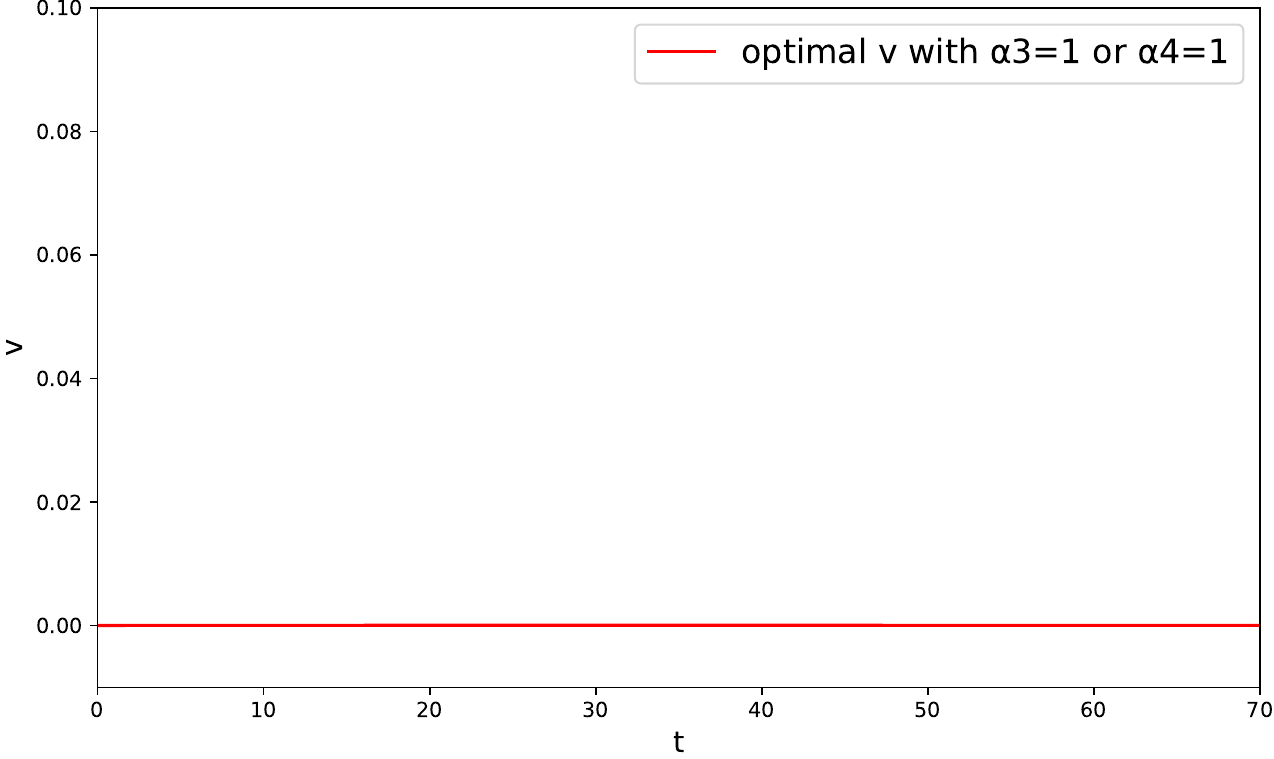} & \includegraphics*[width=0.4\linewidth]{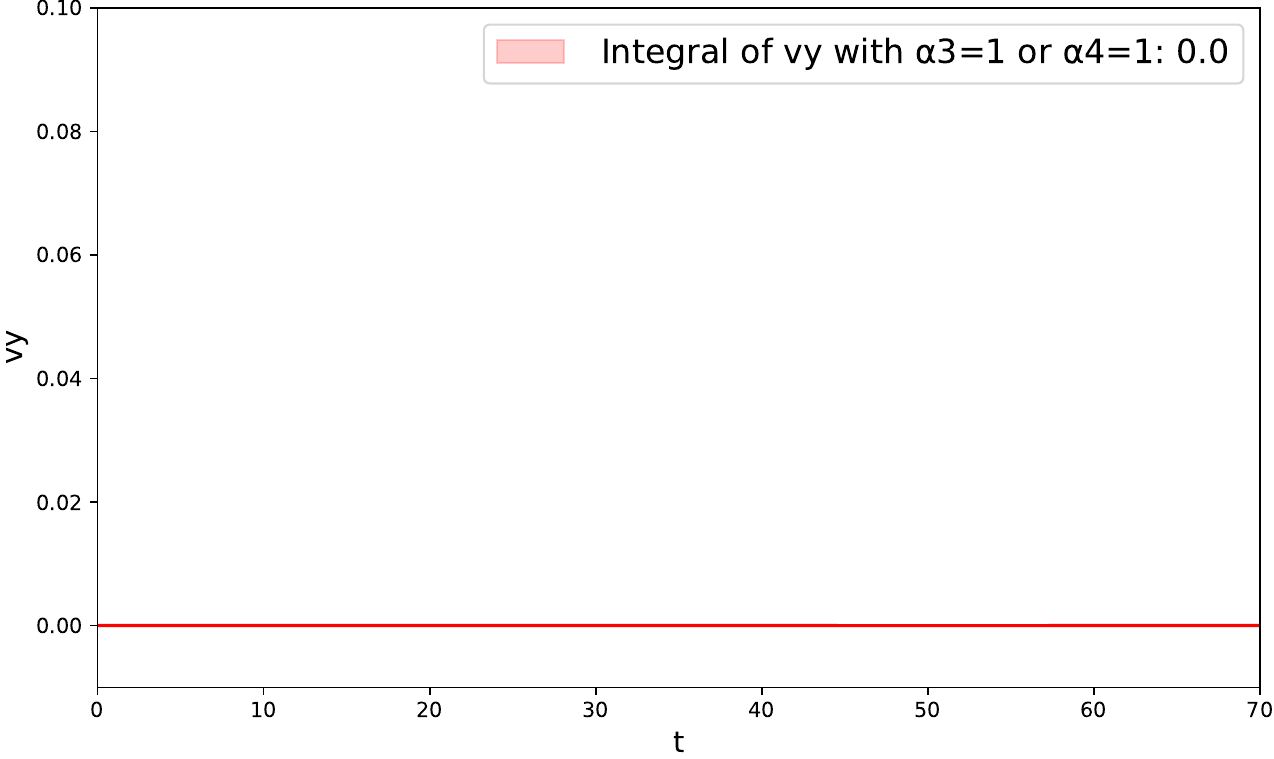} \\
     b) & c)
    \end{tabular} 
    \end{center}
    \caption{{\bf Importance of $\alpha _3$ or $\alpha_4$.} Result obtained by simulating from 0 to 70 time units with 600 steps, minimizing the control effort or the amount of sanctioned individuals, with starting state $w_0=[0.2,0.7,0.1]^T$. a) State space and trajectory of the system. b) Control effort $v$ over time. c) Proportion of punished individuals $yv$ over time. The red shade represents the area under the curve which in this case is zero.}
    \label{fig:a3}
\end{figure}

A similar result is obtained when analyzing the importance of $\alpha_4$. This is because minimizing the frequency of sanctioned individuals implies avoiding sanctioning any of them (\textit{i.e.}, keeping $v$ at zero for the entire period, as with the case of $\alpha_3$). Figure \ref{fig:a3} shows these results%
\footnote{The numerical system trajectory in Figure \ref{fig:a3} a) presents a discrepancy with respect to the closed system trajectory of the free punishment system (see \cite{g12010017}: Figure 1(a)). This is due to the numerical integrator implemented \cite{beal2018gekko}.}%
. Although the single choice of the terms $\alpha_3$ or $\alpha_4$ leads to the no punishment system behavior, they become necessary for combining with a term containing the state error, such as $\alpha_2$, to achieve a trade-off between minimizing the state error and applying punishment.  This will be analyzed in forthcoming sections.

    %\FloatBarrier
      
\subsection{Relative importance of $\alpha_3$ with respect to $\alpha_2$}

When aiming to reduce costs while enhancing collaboration, it is attractive to explore using a combination of $\alpha_2$ and $\alpha_3$ in the cost function (\ref{cost:func}). This approach enables us to balance minimizing the state error trajectory with penalizing only a necessary fraction of free-riders.

To explore the relationship between $\alpha_2$ and $\alpha_3$, we initially assume that $\alpha_2$ is much greater than $\alpha_3$. Afterwards, we decrease the value of $\alpha_2$, while ensuring that $\alpha_2 + \alpha_3 = 1$ for all cases. In the corresponding figures, we use the following color sequence: red, blue, green, cyan, and magenta, which aligns with the mentioned procedure. This process is repeated in the subsequent sections.

\begin{figure}[h!]
    \begin{center}
    %Primeros resultados
    \begin{tabular}{ccc}
    \includegraphics*[width=0.6\linewidth]{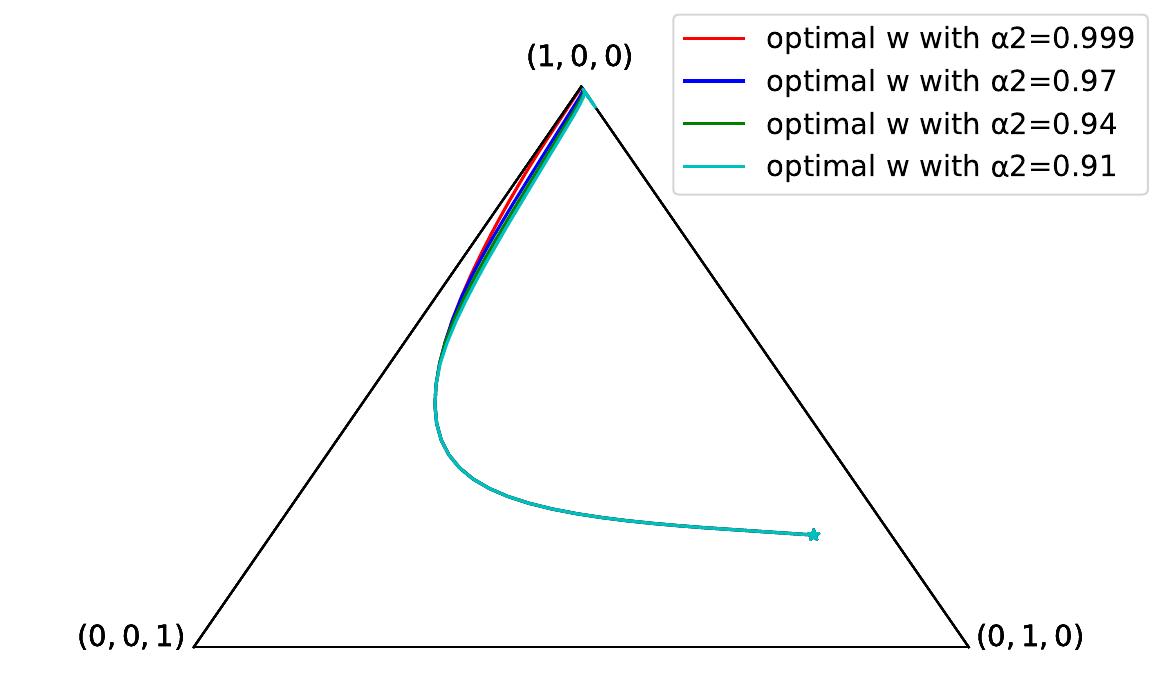} 
    \\ a)
    \end{tabular}
    \begin{tabular}{ccc}
    \includegraphics*[width=0.4\linewidth]{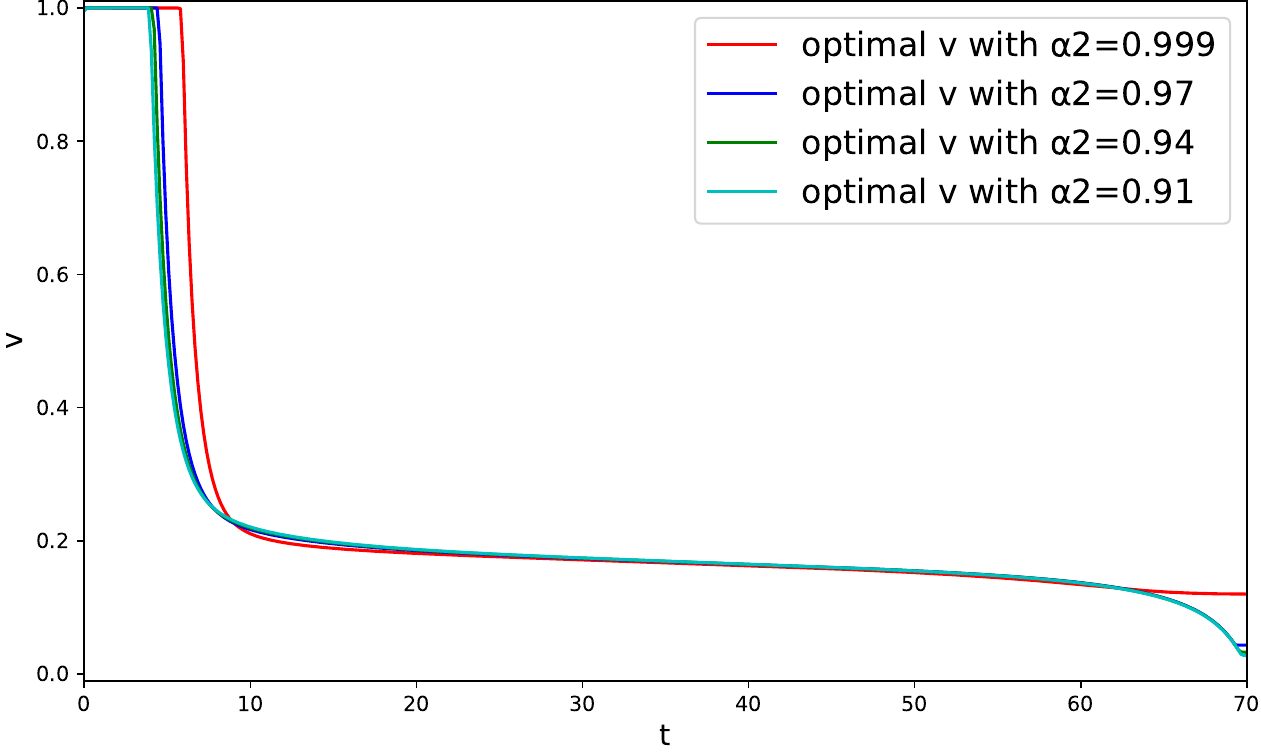} & \includegraphics*[width=0.4\linewidth]{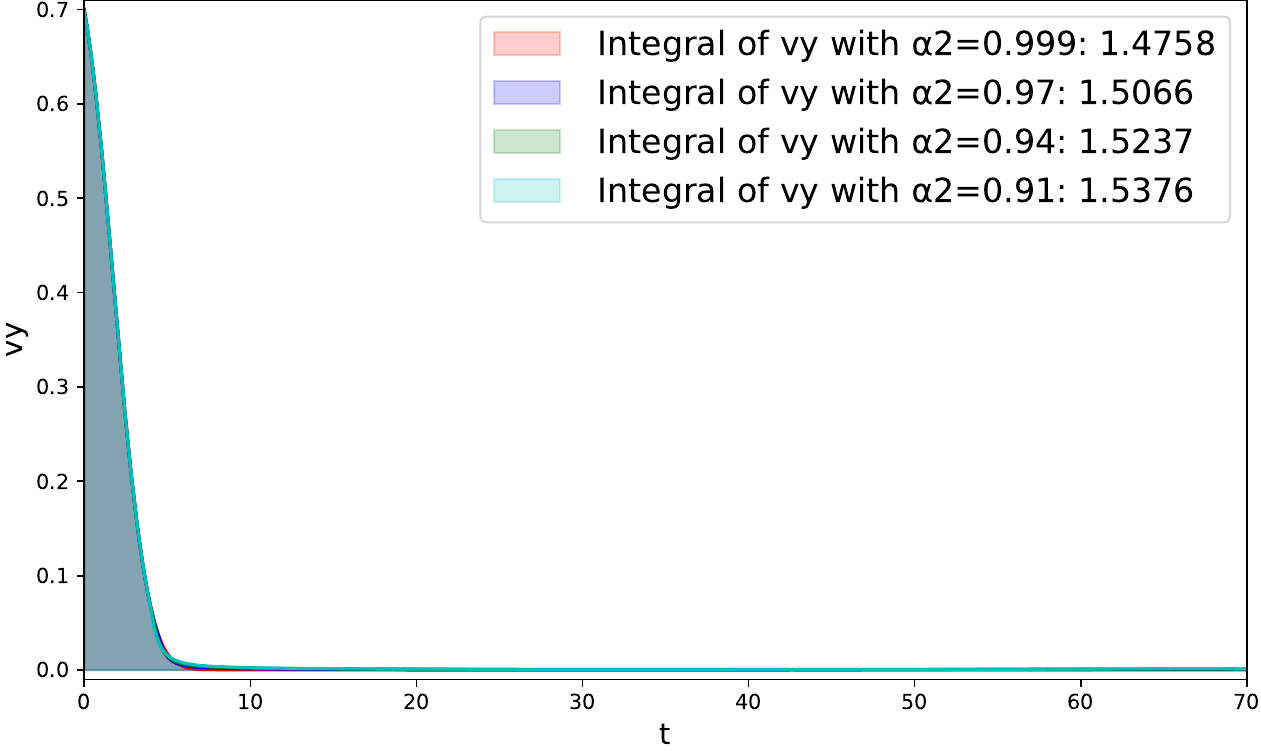} \\
     b) & c)
    \end{tabular}
    \end{center}
    \caption{{\bf Relative importance of $\alpha_3$ with respect to $\alpha _2$. Part 1.} Result obtained by simulating from 0 to 70 time units with 400 steps, such that $\alpha_2$ is 0.999, 0.97, 0.94 and 0.91 and with starting state $w_0=[0.2,0.7,0.1]^T$. a) State space and trajectory of the system. b) Control effort $v$ over time. c) Proportion of punished individuals $yv$ over time. The corresponding shade represents the area under the curve and the legend shows its value.}
    \label{fig:a2a3-1}
\end{figure}

When $\alpha_2$ is much more important than $\alpha_3$ (observe in Figure \ref{fig:a2a3-1} a) and b)) full cooperation is achieved. This is obtained by applying an important amount of punishment at the beginning and maintaining a small amount of punishment at the end.

\begin{figure}[h!]
    \begin{center}
    %Segundos resultados
    \begin{tabular}{ccc}
    \includegraphics*[width=0.6\linewidth]{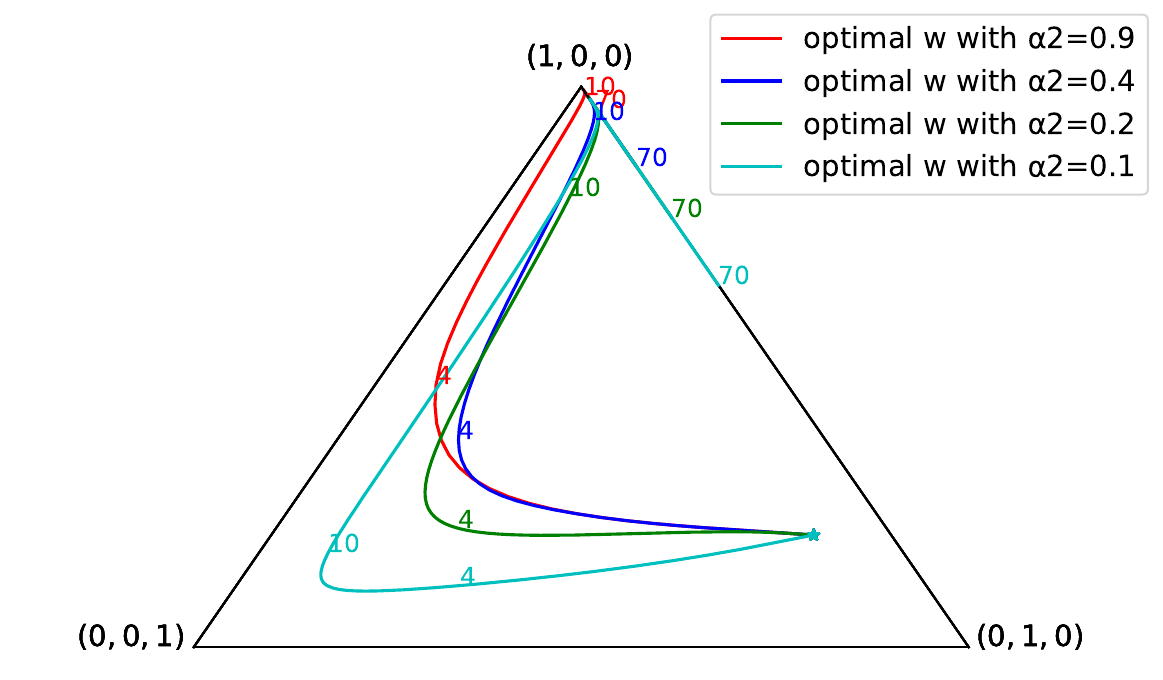} 
    \\ a)
    \end{tabular}
    \begin{tabular}{ccc}
    \includegraphics*[width=0.4\linewidth]{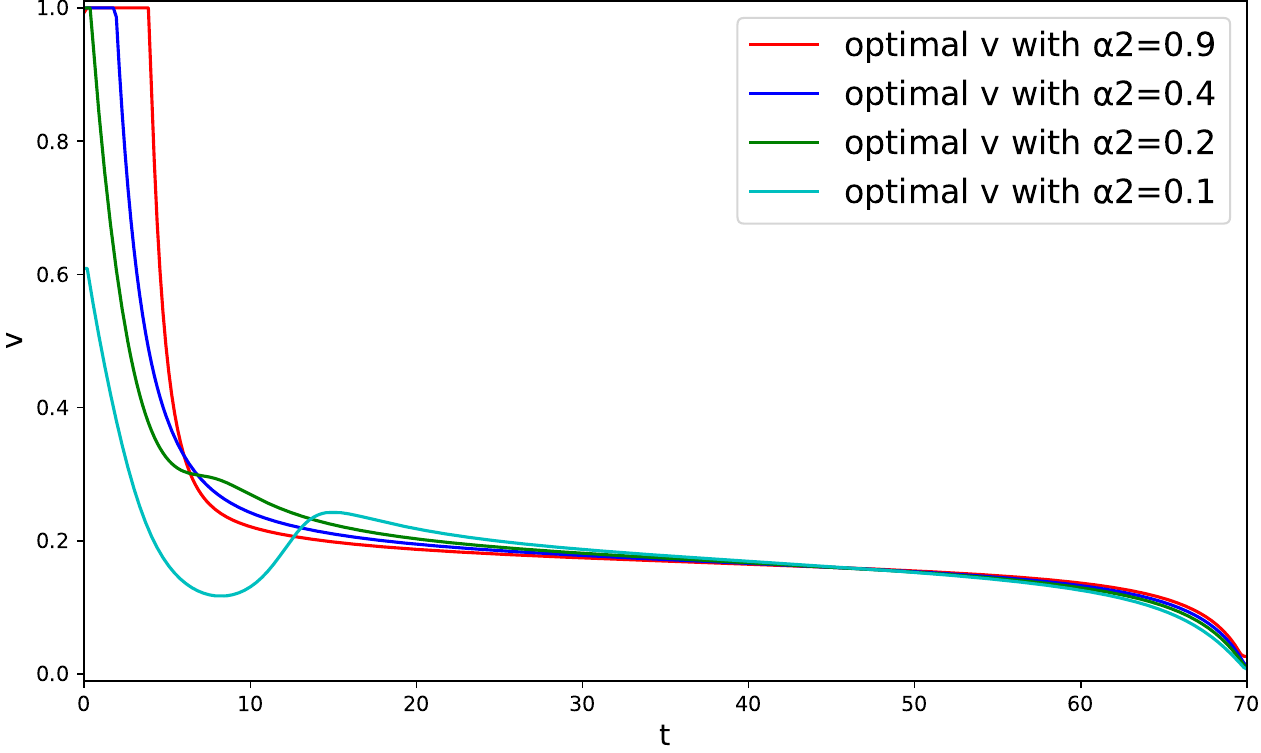} & \includegraphics*[width=0.4\linewidth]{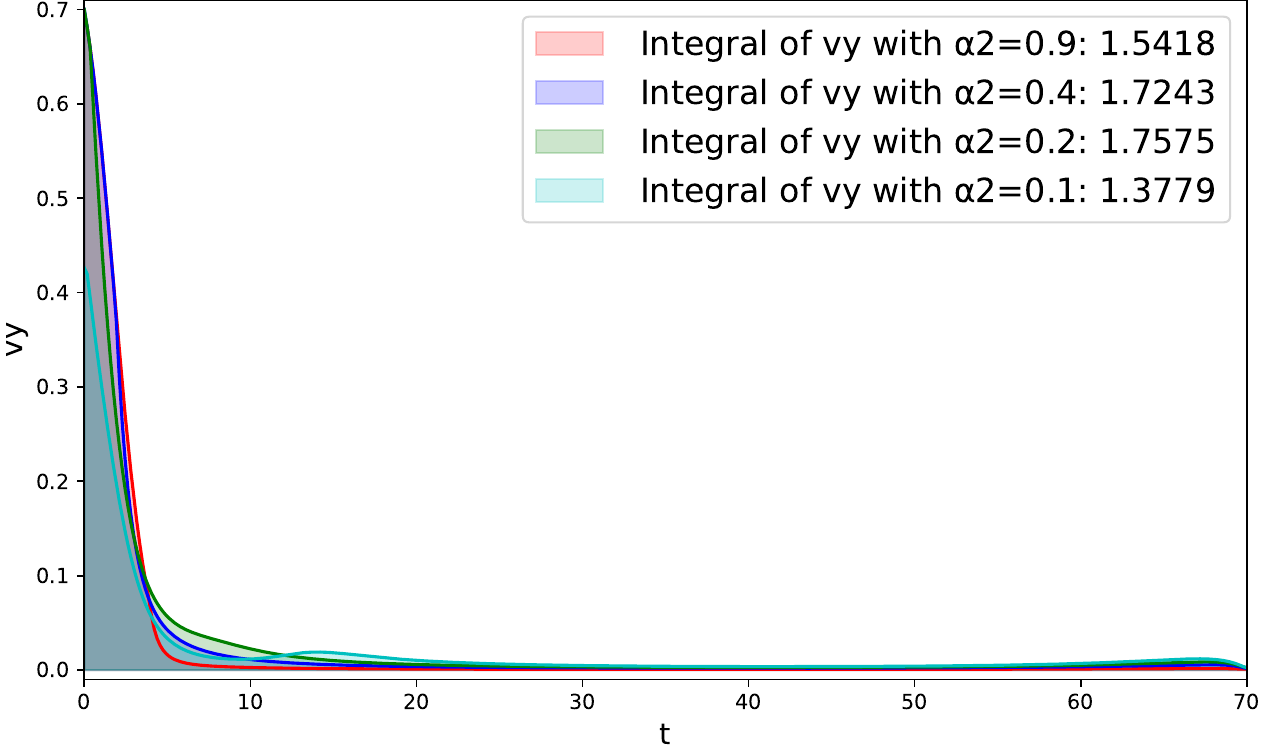} \\
     b) & c)
    \end{tabular}
    \end{center}
    \caption{{\bf Relative importance of $\alpha_3$ with respect to $\alpha _2$. Part 2.} Result obtained by simulating from 0 to 70 time units with 400 steps, such that $\alpha_2$ is 0.9, 0.4, 0.2 and 0.1 with starting state $w_0=[0.2,0.7,0.1]^T$. a) State space and trajectory of the system with timestamps in each corresponding color. b) Control effort $v$ over time. c) Proportion of punished individuals $yv$ over time. The corresponding shade represents the area under the curve and the legend shows its value.}
    \label{fig:a2a3-2}
\end{figure}

As $\alpha_2$ becomes smaller with respect to $\alpha_3$ (as seen in Figure \ref{fig:a2a3-2} a) and b)) a boundary layer emerges in the state trajectory and the controller. Consequently, depending on the pre-specified final error accuracy, the state trajectory can be undesirable for obtaining full cooperation.

%\FloatBarrier
\FloatBarrier

\subsection{Relative importance of $\alpha_4$ with respect to $\alpha_2$} \label{subS_a2a4}

In the real world, another important cost-related factor is the number of punished individuals. To minimize this, we consider the terms associated with the values of $\alpha_2$ and $\alpha_4$ in the cost function (\ref{cost:func}). We use a similar methodology to the previous section, maintaining the same color ordering and ensuring that $\alpha_2+\alpha_4=1$ while reducing the value of $\alpha_2$.

\begin{figure}[h!]
    \begin{center}
    %Primeros resultados
    \begin{tabular}{ccc}
    \includegraphics*[width=0.6\linewidth]{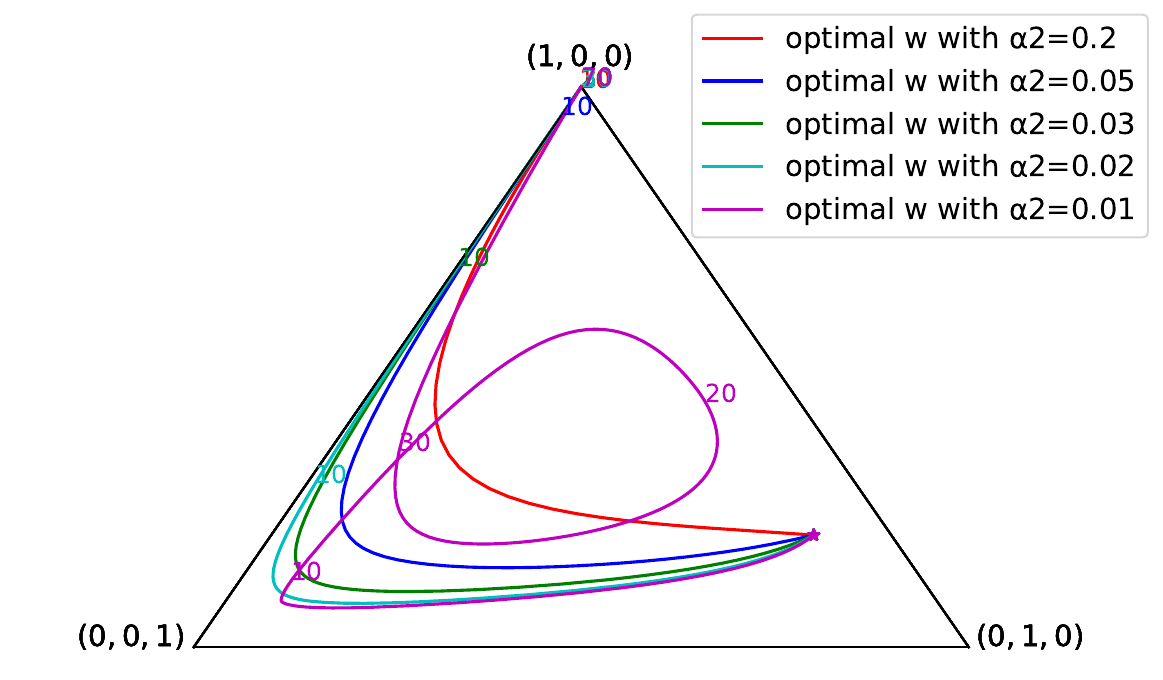} 
    \\ a)
    \end{tabular}
    \begin{tabular}{ccc}
    \includegraphics*[width=0.4\linewidth]{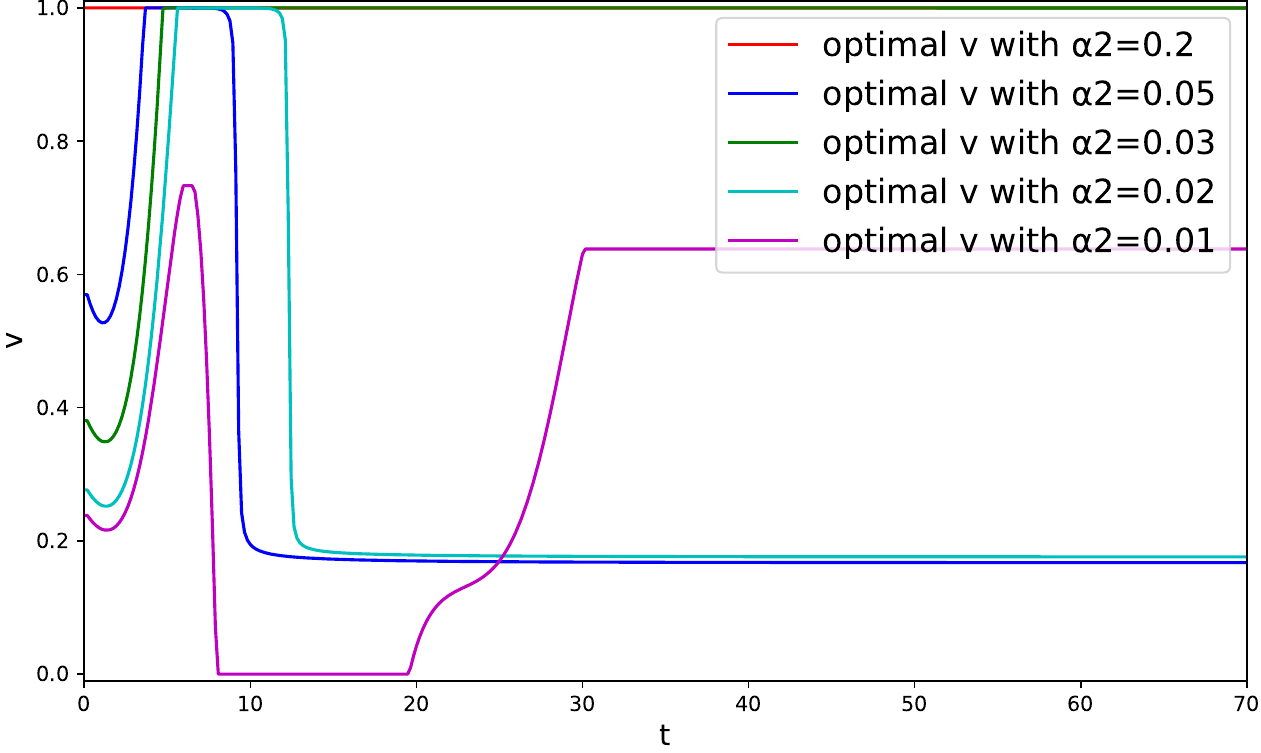} & \includegraphics*[width=0.4\linewidth]{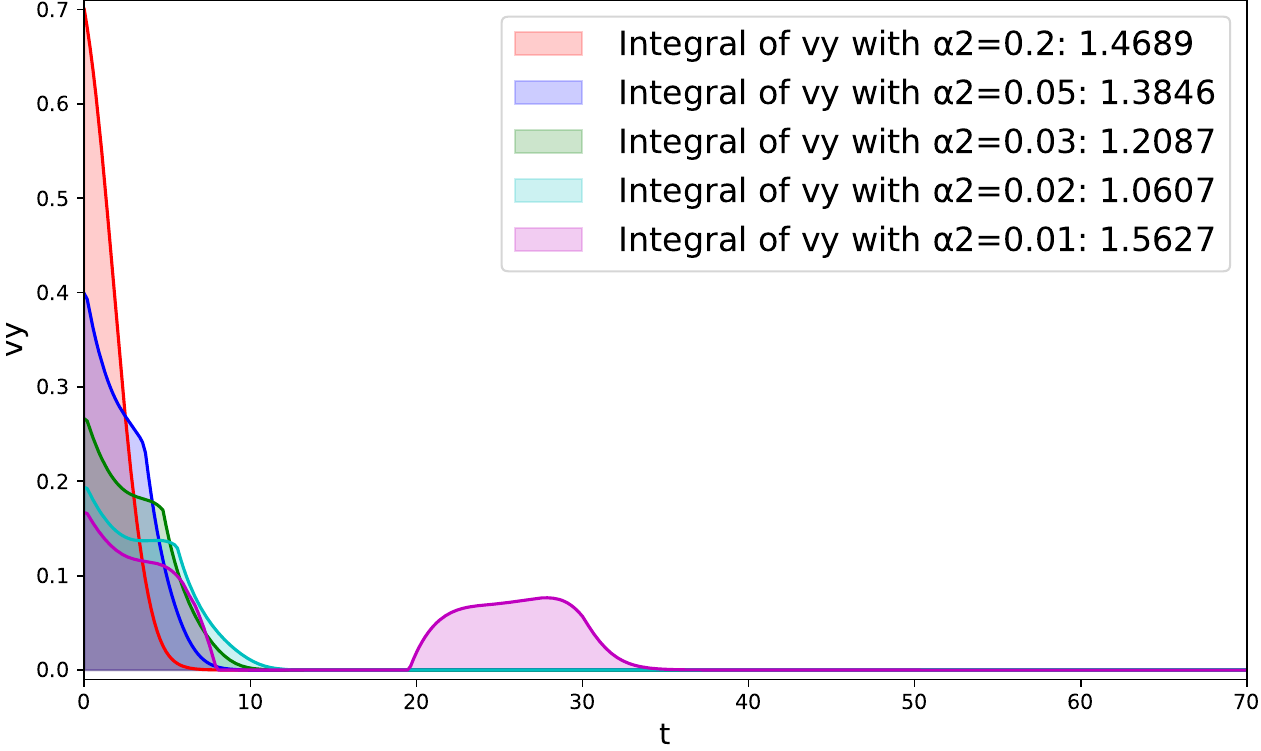} \\
     b) & c)
    \end{tabular}
    \end{center}
        \caption{{\bf Relative importance of $\alpha_4$ with respect to $\alpha_2$. Part 1.} Result obtained by simulating from 0 to 70 time units with 400 steps, such that $\alpha_2$ is 0.2, 0.05, 0.03, 0.02 and 0.01 (with timestamps for this last case) and with starting state $w_0=[0.2,0.7,0.1]^T$. a) State space and trajectory of the system. b) Control effort $v$ over time. c) Proportion of punished individuals $yv$ over time. The corresponding shade represents the area under the curve and the legend shows its value.}
    \label{fig:a2a4-1}
\end{figure}

\begin{figure}[h!]
    \begin{center}
    %Segundos resultados
    \begin{tabular}{ccc}
    \includegraphics*[width=0.6\linewidth]{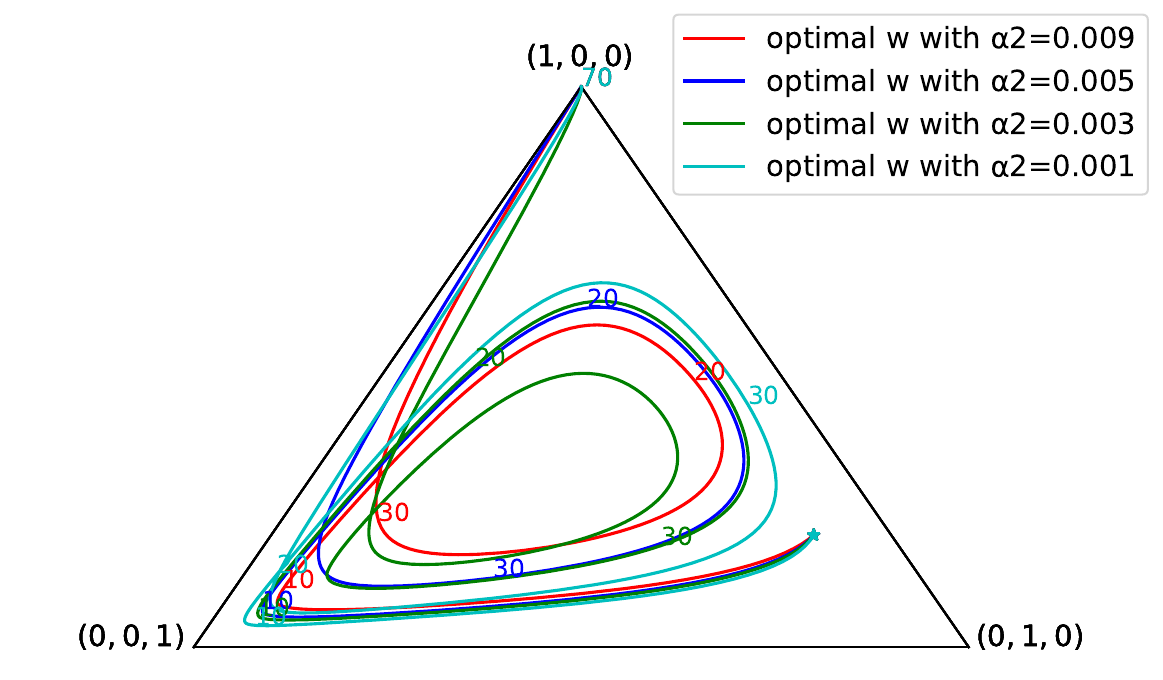} 
    \\ a)
    \end{tabular}
    \begin{tabular}{ccc}
    \includegraphics*[width=0.4\linewidth]{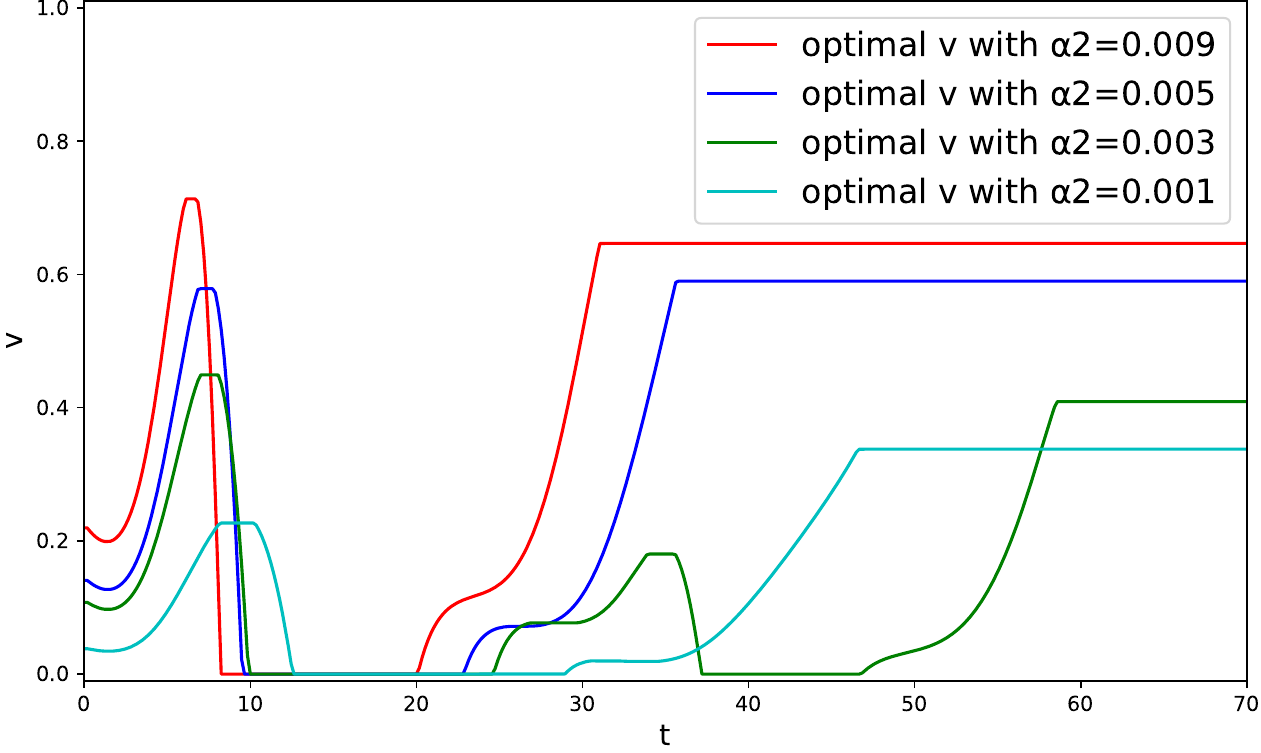} & \includegraphics*[width=0.4\linewidth]{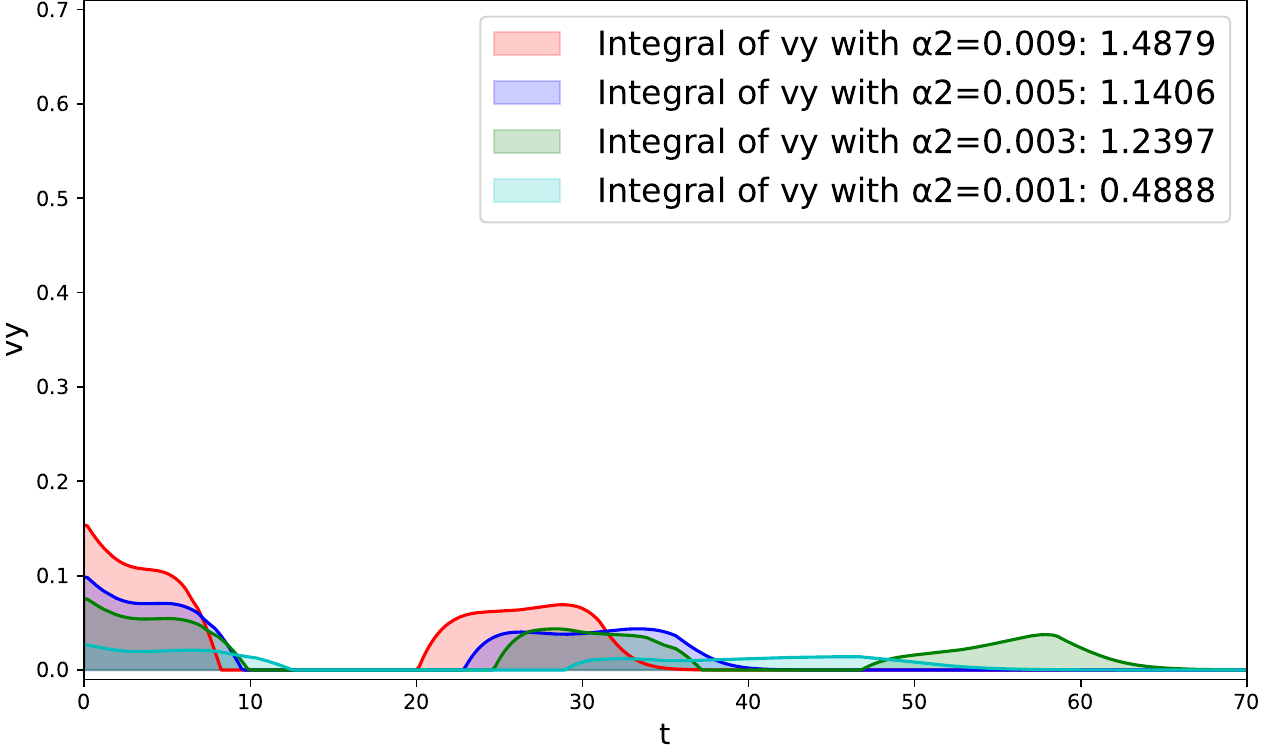} \\
     b) & c)
    \end{tabular}
    \end{center}
    \caption{{\bf Relative importance of $\alpha_4$ with respect to $\alpha_2$. Part 2.} Result obtained by simulating from 0 to 70 time units with 400 steps, such that $\alpha_2$ is 0.009, 0.005, 0.003 and 0.001 and with starting state $w_0=[0.2,0.7,0.1]^T$. a) State space and trajectory of the system. b) Control effort $v$ over time. c) Proportion of punished individuals $yv$ over time. The corresponding shade represents the area under the curve and the legend shows its value.}
    \label{fig:a2a4-2}
\end{figure}

Figure \ref{fig:a2a4-1} shows that when the value of $\alpha_4$ is progressively increased, it reduces both the initial impulse of the controller (Figure \ref{fig:a2a4-1} b)) and the number of punished individuals (Figure \ref{fig:a2a4-1} c)). This provides a counter-intuitive punishment strategy, since in a phase of high defection, the controller punishes fewer individuals. On the contrary, when the value of $\alpha_4$ is decreased, it generates higher values for both the control effort and the number of punished individuals (see Figures \ref{fig:a2a4-2} b) and c)). Still, after a certain minimum threshold value of $\alpha_4$,  the state trajectory becomes more oscillatory in the simplex (see Figures \ref{fig:a2a4-1} a) and \ref{fig:a2a4-2} a)).

The results show that to achieve full cooperation, we can allow for an oscillatory state trajectory by sanctioning a very small number of individuals (represented by the integral of $yv$). However, in practice, this can lead to a high level of desertion and make the institution susceptible to bankruptcy.  In general, even if we do not consider oscillatory trajectories, the results reveal an unexpected effect: we can actually improve cooperation by reducing the number of sanctioned individuals during periods of high free-riding and increasing it during periods of more cooperation.

%Aca tiene que ir esta imagen
\FloatBarrier 

\subsection{Combining the effects of $\alpha_2$, $\alpha_3$ and $\alpha_4$} \label{subS_a2a3a4}

By leveraging the obtained relations between $\alpha_2$ and $\alpha_3$, and $\alpha_2$ and $\alpha_4$, we design a controller that ensures full cooperation while also considering the costs of imposing sanctions during periods of high defection. This approach minimizes the control effort to the optimal level required to sustain cooperation.

Similarly to previous cases, we begin with a low value of $\alpha_2$ while maintaining $\alpha_3=0.0001$ and $\alpha_2+\alpha_3+\alpha_4=1$. Afterward, we increase the relative importance of the term associated with $\alpha_2$ while decreasing the importance of the amount of sanctioned individuals~($\alpha_4$).

\begin{figure}[h!]
    \begin{center}
    %Primeros resultados
    \begin{tabular}{ccc}
    \includegraphics*[width=0.6\linewidth]{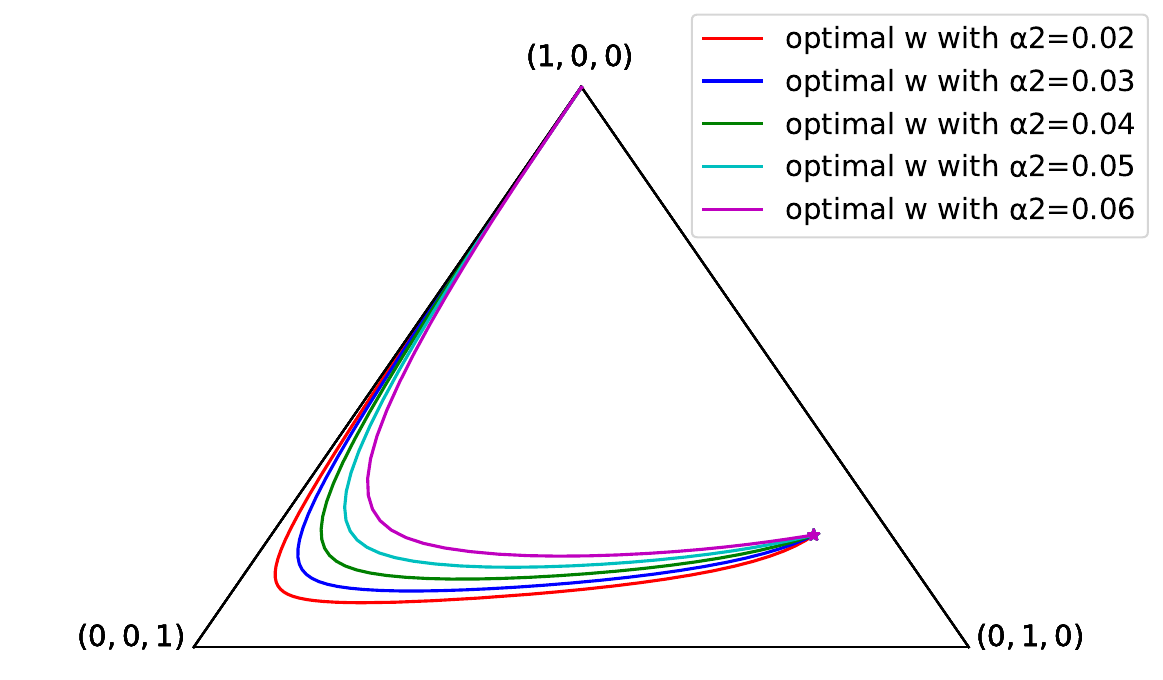} 
    \\ a)
    \end{tabular}
    \begin{tabular}{ccc}
    \includegraphics*[width=0.4\linewidth]{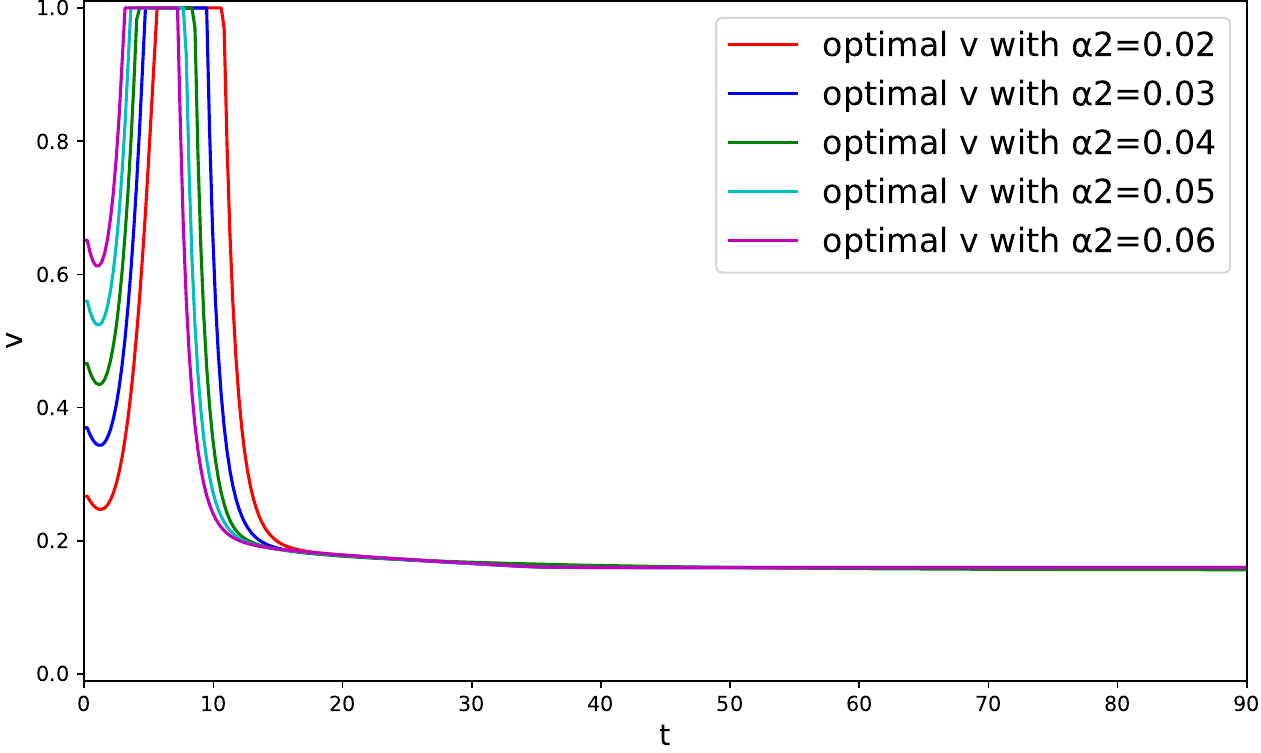} & 
    \includegraphics*[width=0.4\linewidth]{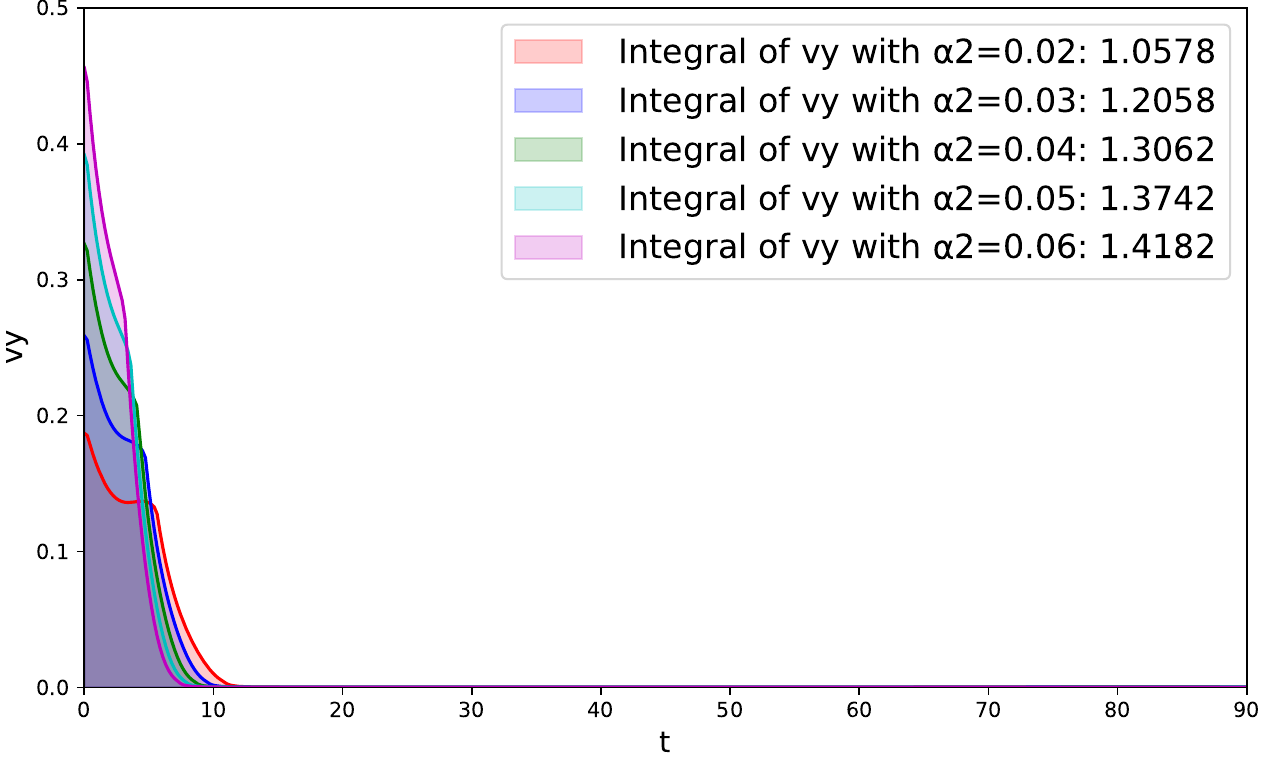} \\
    b) & c)
    \end{tabular}
    \end{center}
    \caption{{\bf Combining results. Part 1.} Result obtained by simulating from 0 to 90 time units with 400 steps, such that $\alpha_2$ is 0.02, 0.03, 0.04, 0.05 and 0.06 and with starting state $w_0=[0.2,0.7,0.1]^T$. a) State space and trajectory of the system. b) Control effort $v$ over time. c) Proportion of punished individuals $yv$ over time. The corresponding shade represents the area under the curve and the legend shows its value.}
    \label{fig:a2a3a4-1}
\end{figure}

Perhaps the most remarkable result is shown in Figure \ref{fig:a2a3a4-1} b). Observe the controller's effort is smaller during high defection periods while being more aggressive when defection has gone down. Finally, it maintains a constant value at the end to discourage defection. The regulation of weights $\alpha_2$ and $\alpha_4$ manifest almost as a time shift of the plateaus in this graph.

Something notable to mention is that when time increases,  the value of $v$ for the test scenarios tends toward the value for which the vertex $x=1$ becomes an attractor. In this case, it is around $v=1/6$ \cite{Botta2020}. Yet, it is not confirmed if this result is just a coincidence given the chosen parameters or if it unveils a special relation of the model.

\subsection{A comparison between punishment strategies}
\label{subS_comp}

In this section, we use a particular scenario (and its associated cost function) to compare finding an optimal solution with using a simple constant value of fractional punishment (or eventually punishing all free-riders at all times). To this end, consider the following cost function:
\begin{eqnarray}
    J_1 =  \int _{0}^{20}\left[
    \frac{0.04}{2} \|w(\tau,v)-w^* \|_2^2+ 
    \frac{0.001}{2} v^2(\tau) +
    \frac{0.959}{2} y^2(\tau)v^2(\tau) 
    \right] d\tau. \label{comp_cost_func}
\end{eqnarray}

\begin{figure}[htb]
    \centering
    \includegraphics*[width=0.5\linewidth]{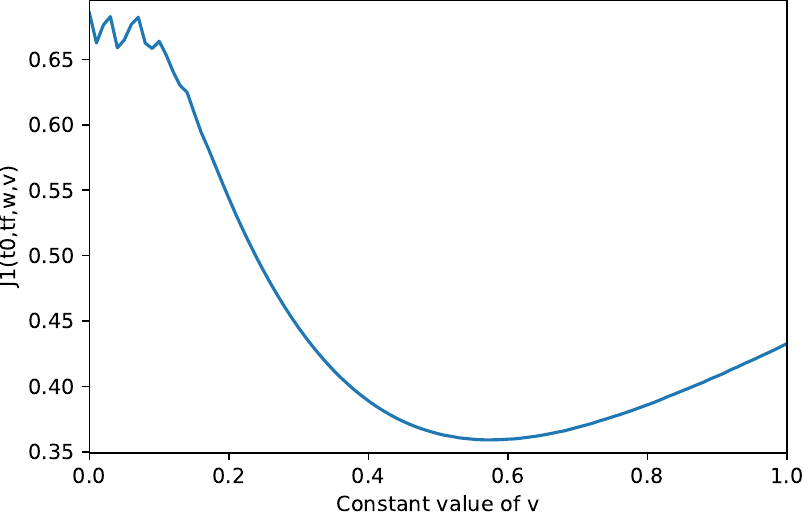}
    \caption{{\bf A comparison between punishment strategies. The cost function when a constant punishment is applied.} Values of cost function (\ref{comp_cost_func}) obtained by taking 101 constant and equally spaced values of $v$ ranging from 0 to 1. Each scenario was tested from initial state $w_0=[0.2,0.7,0.1]^T$ using 20 time units and 400 time steps. From the tests, there is a clear implication that there is a minimum value of $J_1$ at $v \approx 0.57$, and small deviations from this value do not result in big changes of $J_1$.}
    \label{fig:J_vcte}
\end{figure}

For initial conditions $w_0=[0.2,0.7,0.1]^T$, each pair of constant fractional punishment and trajectory $(v,w(v))$, has an associated cost given by $J_1(t_0:=0,t_f:=20,v,w)$. Figure \ref{fig:J_vcte} shows a curve of values of the cost function  (\ref{comp_cost_func}) for several values of~$v$.

%\FloatBarrier

\begin{figure}[h!]
    \begin{center}
    \begin{tabular}{c}
    \includegraphics*[width=0.6\linewidth]{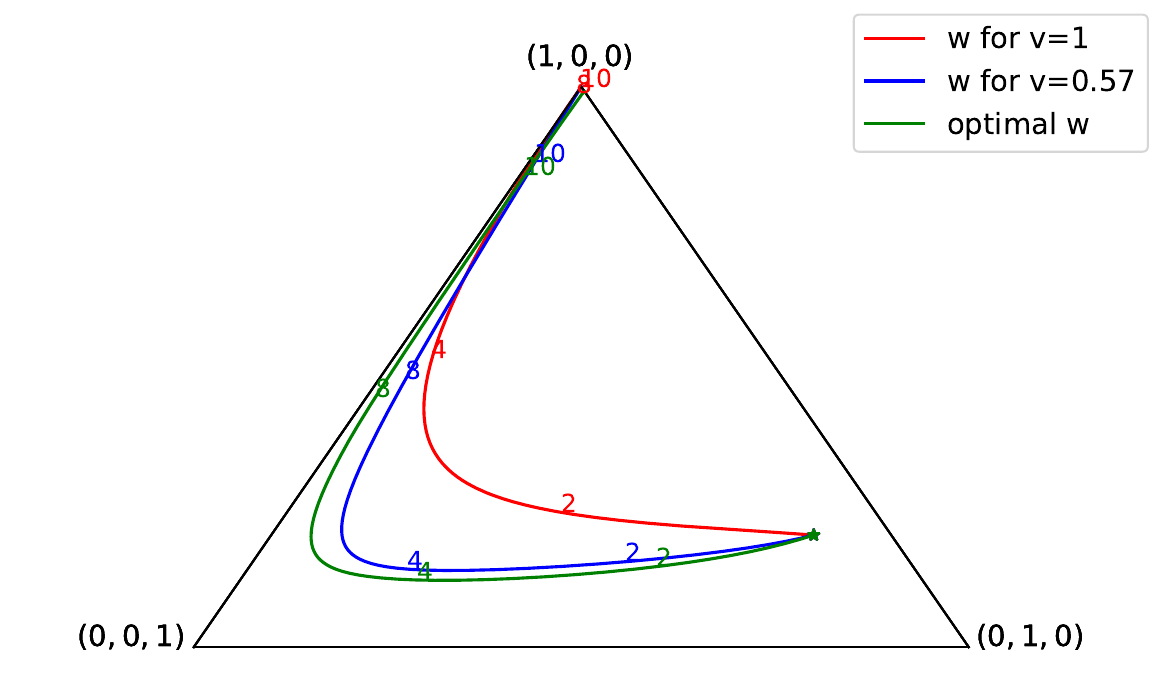}
    \\ a)
    \end{tabular}
    \begin{tabular}{cc}
    \includegraphics*[width=0.4\linewidth]{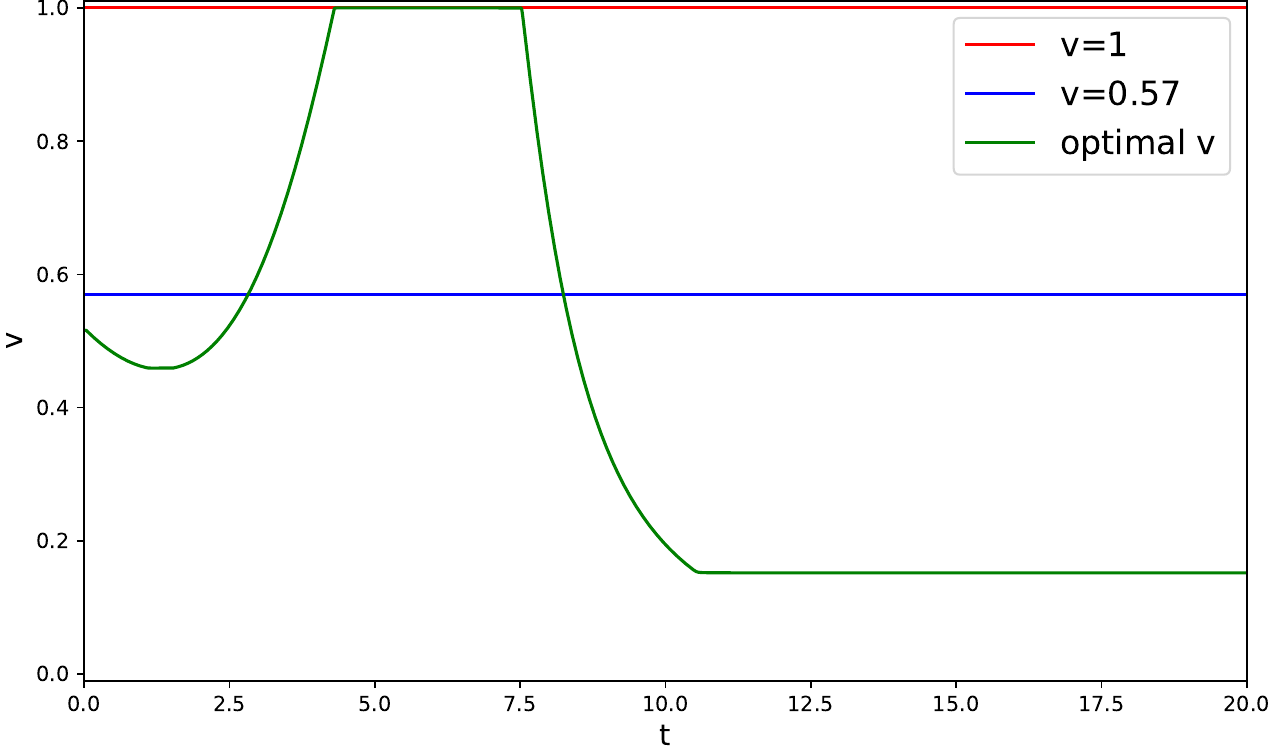} & \includegraphics*[width=0.4\linewidth]{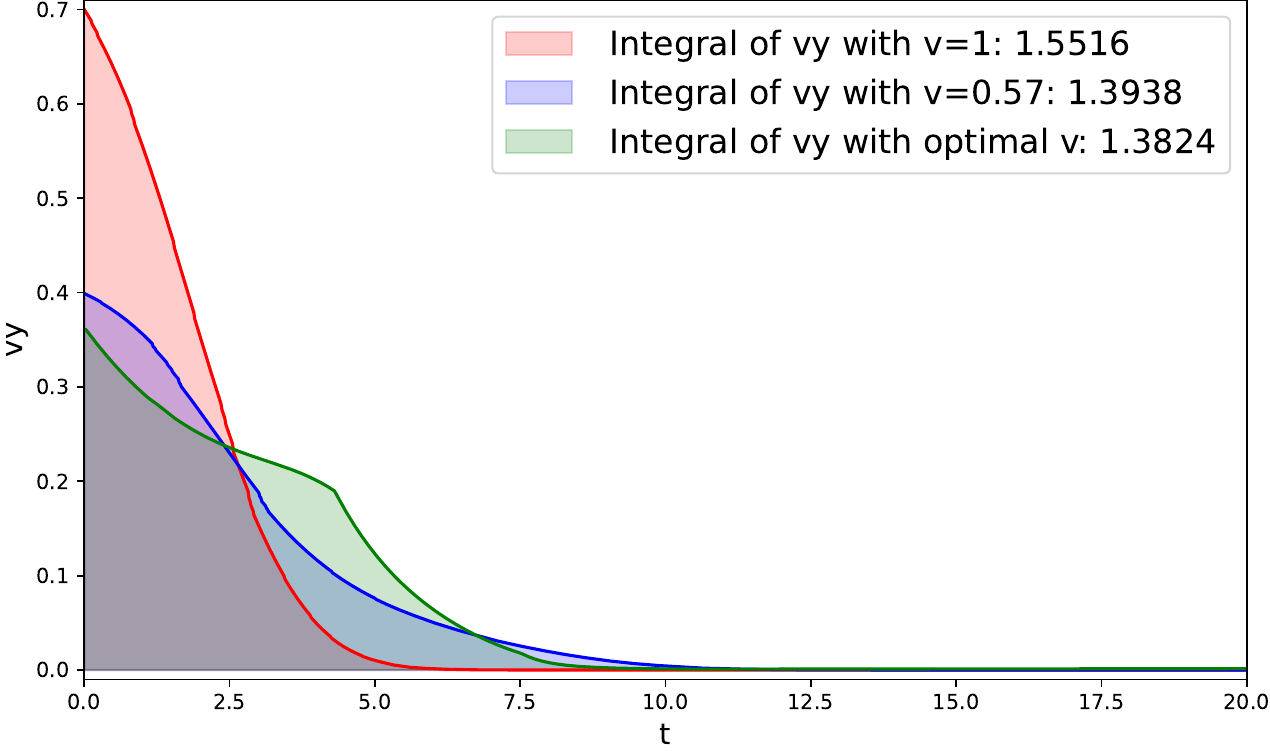} \\
     b) & c)
    \end{tabular}
    \end{center}
    \caption{{\bf A comparison between punishment strategies. Constant fractional punishments against the optimal one.} Result obtained by simulating from 0 to 20 time units with 1200 steps, starting state $w_0=[0.2,0.7,0.1]^T$ and using $v=1$ (in red), $v=0.57$ (in blue) and an optimal $v$ (in green) as punishment strategies. a) State space and trajectory of the system with some color coded time stamps. b) Control effort $v$ over time. c) Proportion of punished individuals $yv$ over time. The corresponding shade represents the area under the curve and the legend shows its value.}
    \label{fig:comp}
\end{figure}

Figure \ref{fig:comp} shows two relevant cases tested against the optimal trajectory obtained using Algorithm \ref{Geko-Alg}. 
It is observed that the optimal solution has a lower value for the cost function and a lower amount of punished individuals (given by the integral of $yv$).
Table \ref{tbl:comp} shows the computational time and the cost function value for some representative values of $v$ and the optimal controller $v(t)$. Notice that obtaining the optimal solution requires less time than obtaining a solution by simulating the 101 cases seen in the tests for Figure \ref{fig:J_vcte}, even when using smaller time steps to find the optimal solution.

%\FloatBarrier

\begin{table}
    \begin{center}
    \caption{Cost comparison between control strategies.}
    \begin{tabular}{|m{3.4cm}|m{3.5cm}|m{2cm}|}%{|c|c|c|}
    \hline 
    Strategy &
    Computation time (using the IPOPT solver)
    & Computed value of $J_1$ \\ \hline
    Constant $v=1$  &  
    \centering  $\leq$ 15s   & 0.4400079 \\
    Constant $v=0.57$  & 
    \centering  $\leq$ 1500s & 0.3638641 \\
    Optimal $v$ &
    \centering  $\leq$ 30s   & 0.3503187 \\
    \hline
    \end{tabular}
    \label{tbl:comp}
    \end{center}
    \footnotesize{\textit{Notes.} This table presents some key features of the comparison shown in Figure \ref{fig:comp}. For this case in particular, the IPOPT solver was used, since even though the APOPT solver has been reliable for most of this work. The IPOPT solver is much faster, even when using many more time steps. For the computation time, for row $v=1$, it just reflects the approximated simulation time; for row $v=0.57$, it mainly reflects the process of finding the minimum for $J_1$ among the 101 values of $v$ mentioned in Figure \ref{fig:J_vcte} and simulating it again with more time steps.} 
\end{table}
\FloatBarrier

\section{Discussion}\label{seccion_discusion}

Trying to account for the most common costs associated with running a public good, three of them naturally arise: the difference between the desired state of cooperation and the actual state of the system (state error), the fraction of free-riders that will be sanctioned and the total amount of people that will be sanctioned. To solve this problem, a weighted cost function that takes into account all this factors is designed, using weights $\alpha_2$, $\alpha_3$ and $\alpha_4$ for the relevant costs mentioned before.

When analyzing each part of the cost function individually, it is observed from the experiments that if any long-term improvement of cooperation is desired, $\alpha_2$ must be considered. Nevertheless, at least one of $\alpha_3$ or $\alpha_4$ should be used to decrease the sanctioning efforts by reducing the fraction of sanctioned individuals (see \S \ref{subS_a1} to \S \ref{subS_a2a4}).

As shown in \S \ref{subS_a2a3a4} a combination of $\alpha_2$, $\alpha_3$, and $\alpha_4$ can be used, such that cooperation improves (because of $\alpha_2$), fractional punishment is low when there are too many free-riders to sanction (because of $\alpha_4$). It is also low when the maintenance of cooperation does not deem it necessary to sanction so many people (because of $\alpha_3$). This is considered a good illustration of the capabilities of the optimal solution obtained for this minimization problem.

Finally in \S \ref{subS_comp} an optimal solution for a specific problem generated from cost function (\ref{comp_cost_func}) is compared with two representative alternative sanctioning strategies: a) sanctioning all free-riders and b) sanctioning a constant fraction of free-riders close to the minimizing value of the cost function.

In summary, this research demonstrates that the optimal solution is better than the best constant fractional punishment in terms of the cost function value and the frequency of punished individuals. It is also found in a considerably shorter time, and the sanctions are applied when they are most effective. These findings clearly show the advantage of the optimal solution over the alternative of finding the best constant fractional punishment.

\section{Conclusion}\label{seccion_conclusion}

The central focus of this paper has been the optimal control of the free-rider problem in an optional public good game using fractional punishment as the controller. Several objective minimization functions were tested based on the error concerning a specific equilibrium point (full cooperation $(x, y, z) =(1, 0, 0)$), the controller's effort, and the frequency of punished free-riders. 

An important term introduced to minimize the cost of punishing is associated with the effort to implement the control given the current frequency of free riders in the population. This term relates to the real per-unit costs of punishing in a public good or service, where the costs depend on the number of punished free-riders.

Numerical experiments evidence that the optimal solutions found result in a lower value of the cost function compared to punishing all or a constant fraction of free-riders at all times.

\bibliographystyle{ieeetr}
\bibliography{Sin_formato}

\begin{thebibliography}{10}

\bibitem{Hamilton1964}
W.~D. Hamilton, ``{The genetical evolution of social behaviour. I.},'' {\em Journal of theoretical biology}, vol.~7, pp.~1--16, July 1964.

\bibitem{Axelrod1981}
R.~Axelrod and W.~Hamilton, ``{The evolution of cooperation},'' {\em Science}, vol.~211, no.~27, pp.~1390 -- 1396, 1981.

\bibitem{ostrom1990governing}
E.~Ostrom, {\em Governing the commons: The evolution of institutions for collective action}.
\newblock Cambridge university press, 1990.

\bibitem{Nowak2006}
M.~A. Nowak, ``{Five Rules for the Evolution of Cooperation},'' {\em Science}, vol.~314, no.~December, 2006.

\bibitem{Sigmund2007}
K.~Sigmund, ``{Punish or perish? Retaliation and collaboration among humans.},'' {\em Trends in ecology \& evolution}, vol.~22, pp.~593--600, Nov. 2007.

\bibitem{Hardin1968}
G.~Hardin, ``{The Tragedy of the Commons},'' {\em Science}, vol.~162, pp.~1243--1248, 1968.

\bibitem{Trivers1971}
R.~L. Trivers, ``{The Evolution of Reciprocal Altruism Published by : The University of Chicago Press Stable URL :},'' {\em The Quarterly Review of Biology}, vol.~46, no.~1, pp.~35--57, 1971.

\bibitem{Yamagishi1986}
T.~Yamagishi, ``{The provision of a sanctioning system as a public good.},'' {\em Journal of Personality and Social Psychology}, vol.~51, no.~1, pp.~110--116, 1986.

\bibitem{Sasaki2012}
T.~Sasaki, A.~Br\"{a}nnstr\"{o}m, U.~Dieckmann, and K.~Sigmund, ``{The take-it-or-leave-it option allows small penalties to overcome social dilemmas.},'' {\em Proceedings of the National Academy of Sciences of the United States of America}, vol.~109, pp.~1165--9, Jan. 2012.

\bibitem{Nowak1992}
M.~Nowak and R.~May, ``{Evolutionary games and spatial chaos},'' {\em Nature}, vol.~359, 1992.

\bibitem{Hauert2002}
C.~Hauert, S.~{De Monte}, J.~Hofbauer, and K.~Sigmund, ``{Volunteering as Red Queen mechanism for cooperation in public goods games.},'' {\em Science (New York, N.Y.)}, vol.~296, pp.~1129--32, May 2002.

\bibitem{Sigmund2010}
K.~Sigmund, C.~Hauert, A.~Traulsen, and H.~Silva, ``{Social Control and the Social Contract: The Emergence of Sanctioning Systems for Collective Action},'' {\em Dynamic Games and Applications}, vol.~1, pp.~149--171, Oct. 2010.

\bibitem{Botta-2024}
R.~Botta, G.~Blanco, and C.~E. Schaerer, ``Discipline and punishment in panoptical public goods games,'' {\em Sci. Rep.}, vol.~14, no.~7903, 2024.

\bibitem{Chen2014}
X.~Chen, A.~Szolnoki, and M.~Perc, ``{Probabilistic sharing solves the problem of costly punishment},'' {\em New Journal of Physics}, vol.~16, 2014.

\bibitem{Dercole2013}
F.~Dercole, M.~{De Carli}, F.~{Della Rossa}, and A.~V. Papadopoulos, ``{Overpunishing is not necessary to fix cooperation in voluntary public goods games.},'' {\em Journal of theoretical biology}, vol.~326, pp.~70--81, jun 2013.

\bibitem{Zhang2017}
J.~Zhang, Y.~Zhu, Z.~Chen, and M.~Cao, ``{Evolutionary Game Dynamics Driven by Setting a Ceiling in Payoffs of Defectors},'' in {\em Proceedings of the 36th Chinese Control Conference July 26-28, 2017, Dalian, China}, pp.~11289--11295, 2017.

\bibitem{g12010017}
R.~Botta, G.~Blanco, and C.~E. Schaerer, ``Fractional punishment of free riders to improve cooperation in optional public good games,'' {\em Games}, vol.~12, no.~1, 2021.

\bibitem{wang2023optimization}
S.~Wang, X.~Chen, Z.~Xiao, A.~Szolnoki, and V.~V. Vasconcelos, ``Optimization of institutional incentives for cooperation in structured populations,'' {\em Journal of the Royal Society Interface}, vol.~20, no.~199, p.~20220653, 2023.

\bibitem{WANG2022_127308}
S.~Wang, X.~Chen, Z.~Xiao, and A.~Szolnoki, ``Decentralized incentives for general well-being in networked public goods game,'' {\em Applied Mathematics and Computation}, vol.~431, p.~127308, 2022.

\bibitem{WANG2019_104914}
S.~Wang, X.~Chen, and A.~Szolnoki, ``Exploring optimal institutional incentives for public cooperation,'' {\em Communications in Nonlinear Science and Numerical Simulation}, vol.~79, p.~104914, 2019.

\bibitem{Hofbauer1998}
J.~Hofbauer and K.~Sigmund, {\em {Evolutionary Games and Population Dynamics}}.
\newblock Cambridge: Cambridge University Press, 1998.

\bibitem{Zeeman1980}
E.~C. Zeeman, ``{Population dynamics from game theory},'' in {\em Global theory of dynamical systems, Lecture Notes in Mathematics}, vol.~819, pp.~471--497, Springer, Berlin, Heidelberg, 1980.

\bibitem{Hauertwz2002}
C.~Hauert, S.~{De Monte}, J.~Hofbauer, and K.~Sigmund, ``{Replicator dynamics for optional public good games},'' {\em Journal of Theoretical Biology}, vol.~218, no.~2, pp.~187--194, 2002.

\bibitem{Hauert2004}
C.~Hauert, N.~Haiden, and K.~Sigmund, ``{The dynamics of public goods},'' {\em Discrete \& Continuous Dynamical Systems - B}, vol.~4, pp.~575--587, may 2004.

\bibitem{Schaerer-Vecpar}
C.~E. Schaerer, T.~Mathew, and M.~Sarkis, ``Block iterative algorithms for the solution of parabolic optimal control problems,'' in {\em High Performance Computing for Computational Science - VECPAR 2006} (M.~Dayd{\'e}, J.~M. L.~M. Palma, {\'A}.~L. G.~A. Coutinho, E.~Pacitti, and J.~C. Lopes, eds.), (Berlin, Heidelberg), pp.~452--465, Springer Berlin Heidelberg, 2007.

\bibitem{Grau-2022}
J.~Grau, R.~Botta, and C.~E. Schaerer, ``Optimal control of fractional punishment in optional public goods game,'' in {\em Proceeding Series of the Brazilian Society of Computational and Applied Mathematics}, vol.~9, (São Carlos/SP), pp.~1--2, SBMAC - Sociedade de Matemática Aplicada e Computacional, 2022.

\bibitem{etna_vol40_pp36-57}
X.~Du, M.~Sarkis, C.~E. Schaerer, and D.~B. Szyld, ``Inexact and truncated parareal-in-time krylov subspace methods for parabolic optimal control problems,'' {\em Electron. Trans. Numer. Anal.}, vol.~40, pp.~36--57, 2013.

\bibitem{doi:10.1137/080717481}
T.~P. Mathew, M.~Sarkis, and C.~E. Schaerer, ``Analysis of block parareal preconditioners for parabolic optimal control problems,'' {\em SIAM Journal on Scientific Computing}, vol.~32, no.~3, pp.~1180--1200, 2010.

\bibitem{Leitao2014}
J.~Baumaister and A.~Leitão, {\em Introdução à Teoria de Controle e Programação Dinâmica}.
\newblock IMPA, 1st. edition~ed., 2014.

\bibitem{Speyer-2010}
J.~L. Speyer and J.~D. H., {\em Primer on Optimal Control Theory}.
\newblock Advances in Design and Control, SIAM, 2010.

\bibitem{beal2018gekko}
L.~Beal, D.~Hill, R.~Martin, and J.~Hedengren, ``Gekko optimization suite,'' {\em Processes}, vol.~6, no.~8, p.~106, 2018.

\bibitem{Botta2020}
R.~Botta, C.~E. Schaerer, and G.~Blanco, ``{Cooperation and punishment in community managed water supply system},'' in {\em Conference of Computationar Interdisciplinary Science, 2019, Atlanta. Anais eletr{\^{o}}nicos. Campinas, Galo{\'{a}}, 2019.}, 2020.

\end{thebibliography}

\end{document}